\documentclass[aps,prd,twocolumn,floatfix,groupedaddress]{revtex4}
\usepackage{epsfig}
\usepackage{dcolumn}   % needed for some tables
\usepackage{bm}   
\usepackage{graphicx}     % for math
\usepackage{amssymb}  
\begin{document}
\title{Dynamics of Chemical Equilibrium of Hadronic Matter Close to $T_c$}

\author{J. Noronha-Hostler$^{1}$}
\author{M. Beitel$^{2}$}
\author{C. Greiner$^{2}$}
\author{I. Shovkovy$^{3}$}
\affiliation{$^{1}$The Frankfurt
International Graduate School for Science (FIGSS), D-60438 Frankfurt am Main, Germany}
\affiliation{$^{2}$Institut f\"ur Theoretische Physik, Johann Wolfgang
Goethe--Universit\"at, D-60438 Frankfurt am Main, Germany}
\affiliation{$^{3}$Department of Applied Sciences and Mathematics, Arizona State University, Mesa, Arizona 85212, USA}
\date{\today}
\begin{abstract}

Quick chemical equilibration times of hadrons (specifically, $p\bar{p}$ , $K\bar{K}$ , $\Lambda\bar{\Lambda}$, and $\Omega\bar{\Omega}$ pairs) within a hadron gas are explained 
dynamically 
using Hagedorn states, which drive particles into equilibrium close to the critical temperature.  Within this scheme, we use master equations and 
derive various analytical estimates for the chemical equilibration times. We compare our model to recent lattice results and find that for both $T_c=176$ MeV and $T_c=196$ MeV, the hadrons can reach chemical equilibrium almost immediately, well before the chemical freeze-out temperatures found in thermal fits for a hadron gas without Hagedorn states. Furthermore the ratios  $p/\pi$, $K/\pi$ , $\Lambda/\pi$,
and $\Omega / \pi $ match experimental values well in our dynamical scenario.

\end{abstract}
%\pacs{12.38.Mh, 25.75.Dw, 24.85.1p}
\maketitle

\section{Introduction}

(Anti-)strangeness enhancement was first observed at CERN-SPS energies by comparing anti-hyperons, multi-strange baryons, and kaons to $pp$-data.  It was considered a signature for quark gluon plasma (QGP) because, using binary strangeness production and exchange reactions, chemical equilibrium could not be reached  within a standard hadron gas phase, i.e., the chemical equilibration time was on the order of 
$\tau\sim 100-1000\frac{\mathrm{fm}}{\mathrm{c}}$ 
whereas the lifetime of a fireball in the hadronic stages is only $\tau\approx
4-7 \frac{\mathrm{fm}}{\mathrm{c}}$ \cite{Koch:1986ud}.  It was then proposed that there exists a strong hint for QGP at SPS because strange quarks can be produced more abundantly by gluon fusion, which would account for strangeness enhancement following hadronization and rescattering of strange quarks. Later, however, multi-mesonic reactions were used to explain secondary production of $\bar{p}$ and anti-hyperons \cite{Rapp:2000gy,Greiner}.
At SPS they give a chemical equilibration time $ \tau_{\bar{Y}}\approx 1-3\frac{\mathrm{fm}}{c}$ using an annihilation cross section of $\sigma_{\rho\bar{Y}}\approx\sigma_{\rho\bar{p}}\approx 50\mathrm{mb}$ and  a baryon density of $\rho_{B}\approx \rho_{0}\;\mathrm{to}\;2\rho_{0}$, which is typical for 
evolving strongly interacting matter at SPS before chemical freeze-out.
Therefore, the time scale is
short enough to account for chemical equilibration within a cooling hadronic
fireball at SPS.

A problem arises when the same multi-mesonic reactions were employed in the hadron gas phase at RHIC temperatures  where experiments again show that the particle abundances reach chemical equilibration close to the phase transition \cite{Braun-Munzinger}.  At RHIC at $T=170$ MeV, where $\sigma\approx
30\mathrm{mb}$ and 
$\rho_{B}^{eq}\approx\rho_{\bar{B}}^{eq}\approx0.04\mathrm{fm}^{-3}$, the equilibrium rate for (anti-)baryon production is
$\tau\approx
10\frac{\mathrm{fm}}{\mathrm{c}}$.  

Moreover, $\tau\approx
10\frac{\mathrm{fm}}{\mathrm{c}}$ was also obtained in Ref.\ \cite{Kapusta} using a fluctuation-dissipation theorem.
From hadron cascades a significant deviation  was found from the chemically saturated strange (anti-)baryons  yields in the $5\%$ most central Au-Au collisions \cite{Huovinen:2003sa}. 
These discrepancies suggest that hadrons are ``born" into equilibrium, i.e.,
the system is already in a chemically frozen out state at the end of
the phase transition \cite{Stock:1999hm,Heinz:2006ur}.  
In order to circumvent such long time scales it was suggested that near $T_{c}$ there exists an extra large particle density overpopulated with pions and kaons, which drive the baryons/anti-baryons into equilibrium \cite{BSW}.  But it is not clear how this overpopulation should appear, and how the subsequent population of (anti-)baryons would follow.   Moreover, the overpopulated (anti-)baryons do not later disappear \cite{Greiner:2004vm}. Therefore, it was conjectured that Hagedorn resonances (heavy resonances near $T_{c} $ with an exponential mass spectrum) could account for the extra (anti-)baryons \cite{Greiner:2004vm,Noronha-Hostler:2007fg,NoronhaHostler:2009hp}.  

Hadrons can develop according to
\begin{eqnarray}\label{eqn:decay}
n\pi&\leftrightarrow &HS\leftrightarrow n^{\prime}\pi+X\bar{X}
\end{eqnarray} 
where $X\bar{X}$ can be substituted with  $p\bar{p}$ , $K\bar{K}$ , $\Lambda\bar{\Lambda}$, or $\Omega\bar{\Omega}$.    
Eq.\ (\ref{eqn:decay}) provides an efficient method for producing of $X\bar{X}$ pairs  because of the large decay widths of the Hagedorn states. In Eq. (\ref{eqn:decay}), $n$ is the number of pions for the decay $n\pi\leftrightarrow HS$ and $n^{\prime}$  is the number of pions that a Hagedorn state will decay into when a $X\bar{X}$ is present.
Since Hagedorn resonances are highly unstable, the phase space for multi-particle decays drastically increases when the mass increases.  Therefore, the resonances catalyze rapid equilibration of $X\bar{X}$ near $T_{c} $ and die out moderately below $T_c$ \cite{Noronha-Hostler:2007fg}.

Unlike in pure glue $SU(3)$ gauge theory where the Polyakov loop is the order parameter for the deconfinement transition (which is weakly first-order), the rapid crossover seen on lattice calculations involving dynamical fermions indicates that there is not a well defined order parameter that can distinguish the confined phase from the deconfined phase. Because of this it is natural to look for a hadronic mechanism for quick chemical equilibration near the phase transition.  One such possibility could be the inclusion of Hagedorn states. Recently, Hagedorn states have been shown to contribute to the physical description of a hadron gas close to $T_c$.  The inclusion of Hagedorn states leads to a low $\eta/s$ in the hadron gas phase \cite{NoronhaHostler:2008ju}, which nears the string theory bound $\eta/s=1/(4\pi)$ \cite{KSS}. Calculations of the trace anomaly including Hagedorn states also fits recent lattice results well and correctly describe the minimum of  the speed of sound squared, $c_s^2,$ near the phase transition found on the lattice \cite{NoronhaHostler:2008ju}. Estimates for the bulk viscosity including Hagedorn states in the hadron gas phase indicate that the bulk viscosity, $\zeta/s$, increases near $T_c$, which agrees with the general analysis done in \cite{Kharzeev:2007wb}. Furthermore, it has been shown \cite{NoronhaHostler:2009tz} that Hagedorn states provide a better fit within a thermal model to the hadron yield particle ratios.  Additionally, Hagedorn states provide a mechanism to relate $T_c$ and $T_{chem}$, which then leads to the suggestion that a lower critical temperature could possibly be preferred, according to the thermal fits \cite{NoronhaHostler:2009tz}. 

Previously, in Ref.\ \cite{Noronha-Hostler:2007fg} we presented analytical results, which we will derive in detail here.  Moreover, we saw that both the baryons and kaons equilibrated quickly within an expanding fireball.  The initial saturation of pions, Hagedorn states, baryons, and kaons played no significant role in the ratios such as $K/\pi$ and $\left(B+\bar{B}\right)/\pi$. 

Here we consider the effects of various initial conditions on the chemical freeze-out temperature and we find that while they play a small role on the total particle number, they still reproduce fast chemical equilibration times.  
Additionally, we assume lattice values of the critical temperatures ($T_c=176$ MeV \cite{zodor} and $T_c=196$ MeV \cite{Cheng:2007jq,Bazavov:2009zn}) and find that chemical equilibrium abundances are still reached close to the temperature given by thermal fits ($T\approx 160$ MeV).

This paper is structured in the following manner. In Section \ref{sec:model} we discuss the details of our statistical model that calculates the chemical equilibrium values of the Hagedorn states and other hadrons.  Furthermore in this section, fits are shown to thermodynamical properties calculated in lattice QCD, which are used to determine the mass spectrum of the Hagedorn states and the rate equations are discussed in detail.  In Section \ref{tau} we are able to extract the chemical equilibration time of an $X\bar{X}$ pair when the pions and Hagedorn states are held constant. In Section \ref{ar} we derive an analytical result of the rate equations when we consider only the decay $HS\leftrightarrow n\pi$.  We then discuss the case of an expanding fireball and the results for the various $X\bar{X}$ pair
production in Section \ref{expansion}. The production of $\Omega $ particles will also be considered
in Section \ref{omega}. 
We  summarized and discussed our results in Section \ref{conclusions}.  
In Appendix \ref{app} we present some analytical and numerical results for the various equilibration stages in the hadron and Hagedorn states gas mixture.

\section{Model}\label{sec:model}

Hagedorn resonances have an exponentially growing mass spectrum \cite{Hagedorn:1968jf}. Their large masses open up the phase space for multi-particle decays.
Recent analysis involving Hagedorn states is given in \cite{Broniowski:2004yh}.  Moreover, thoughts on observing Hagedorn states in experiments are given in \cite{Bugaev:2008nu} and their usage as a thermostat in \cite{Moretto:2006zz}. 
Hagedorn states can also explain the phase transition
{\em above } the critical temperature and, depending on the intrinsic parameters, 
the order of the phase transition \cite{Zakout:2006zj}.
For the following discussion, the  
overall density of
Hagedorn states in our extended Hagedorn gas model are straightforwardly described by, 
\begin{equation}\label{eqn:fitrho}
    \rho=\int_{M_{0}}^{M}\frac{A}{\left[m^2 +m_{r}^2\right]^{\frac{5}{4}}}e^{\frac{m}{T_{H}}}dm.
\end{equation}
where $M_{0}=2$ GeV and $m_{r}^2=0.5$ GeV. 
We note that in this work we consider only mesonic Hagedorn states with no net strangeness.
The exponential in Eq.\ (\ref{eqn:fitrho}) arises from Hagedorn's original idea that there is an exponentially growing mass spectrum. Thus, as $T_H$ is approached, Hagedorn states become increasingly more relevant and heavier resonances ``appear".  The factor in front of the exponential has various forms \cite{Broniowski:2004yh,Moretto:2006zz}. While the choice in this factor can vary, it was found in \cite{Broniowski:2004yh} that the present form gives lower values of $T_H$, which more closely match the predicted lattice critical temperature \cite{Cheng:2007jq,Bazavov:2009zn,zodor}.

Returning to Eq.\ (\ref{eqn:fitrho}), its parameters (A, M, and $T_H$) are dependent on the critical temperature. We assume that $T_H=T_c$, and then we consider the two different different lattice results for $T_c$: $T_c=196$ MeV \cite{Cheng:2007jq,Bazavov:2009zn}, which uses an almost physical pion mass, and $T_c=176$ MeV \cite{zodor}. Furthermore, we need to take into account the repulsive interactions and, therefore, we use the following volume corrections (as also seen in \cite{NoronhaHostler:2008ju,Kapusta:1982qd,Rischke:1991ke}):
\begin{eqnarray}\label{eqn:cor}
T&=&\frac{T^*}{1-\frac{p_{pt}\left(T^*,\;\mu_b^*\right)}{4B}}\nonumber\\
\mu_b&=&\frac{\mu_b^*}{1-\frac{p_{pt}\left(T^*,\;\mu_b^*\right)}{4B}}\nonumber\\
p_{xv}&=&\frac{p_{pt}\left(T^*,\mu_b^*\right)}{1-\frac{p_{pt}\left(T^*,\;\mu_b^*\right)}{4B}}\nonumber\\
\varepsilon_{xv}&=&\frac{\varepsilon_{pt}\left(T^*,\mu_b^*\right)}{1+\frac{\varepsilon_{pt}\left(T^*,\;\mu_b^*\right)}{4B}}\nonumber\\
n_{xv}&=&\frac{n_{pt}\left(T^*,\mu_b^*\right)}{1+\frac{\varepsilon_{pt}\left(T^*,\;\mu_b^*\right)}{4B}},\nonumber\\
s_{xv}&=&\frac{s_{pt}\left(T^*,\mu_b^*\right)}{1+\frac{\varepsilon_{pt}\left(T^*,\;\mu_b^*\right)}{4B}},
\end{eqnarray}
which ensure that the our model is thermodynamically consistent. Note that $B$ is a free parameter that is based upon the idea of the MIT bag constant.  

\begin{figure}[h]
\centering
\includegraphics[width=3.in]{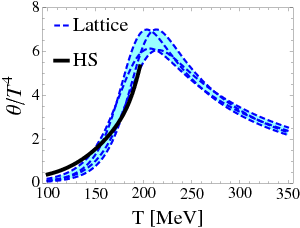}
\caption{Comparison of trace anomaly to lattice QCD results from  \cite{Cheng:2007jq,Bazavov:2009zn} where $T_c=196$ MeV.   }
\label{fig:eT4}
\end{figure}

\begin{figure}[h]
\centering
\includegraphics[width=3.in]{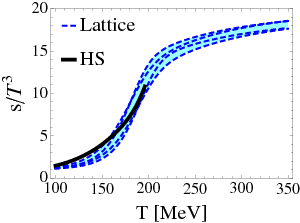}
\caption{Comparison of entropy density to lattice QCD results from  \cite{Cheng:2007jq,Bazavov:2009zn} where $T_c=196$ MeV.   }
\label{fig:sT4}
\end{figure}

In order to find the maximum Hagedorn state mass $M$ and the ``degeneracy" A, we fit our model to the thermodynamic properties of the lattice.  
In the RBC-Bielefeld collaboration the thermodynamical properties are derived from 
the quantity $\varepsilon-3p$, the so called
interaction measure, which is what we fit in order to obtain the parameters for the Hagedorn states.  Thus, we obtain $T_H=196$ MeV, $A=0.5 GeV^{3/2}$, $M=12$ GeV, and $B=\left(340 GeV\right)^4$.  The fit for the trace anomaly $\Theta/T^4$ is shown in Fig.\ \ref{fig:eT4}.  We also show the fit for the entropy density in Fig.\ \ref{fig:sT4}.  Both fits are within the error of lattice and mimic the behavior of the lattice results.  As discussed in \cite{NoronhaHostler:2008ju}, a hadron resonance gas model with Hagedorn states uniquely fits the lattice data whereas a hadron resonance gas without Hagedorn states completely misses the behavior.

\begin{figure}[h]
\centering
\includegraphics[width=3.in]{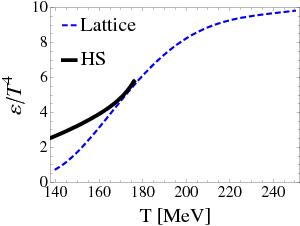}
\caption{Comparison of energy density to lattice QCD results from  \cite{zodor} where $T_c=176$ MeV.   }
\label{fig:eT4_173}
\end{figure}

\begin{figure}[h]
\centering
\includegraphics[width=3.in]{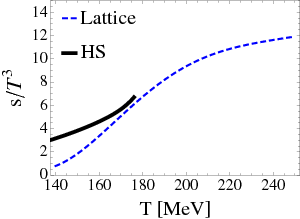}
\caption{Comparison of entropy density to lattice QCD results from \cite{zodor} where $T_c=176$ MeV. }
\label{fig:sT4_173}
\end{figure}

BMW calculates the thermodynamical properties separately and, therefore, we fit only the energy density as shown in Fig.\ \ref{fig:eT4_173}.  From that we obtain $T_H=176$ MeV, $A=0.1 GeV^{3/2}$, $M=12$ GeV, and $B=\left(300 GeV\right)^4$. We also show a comparison to the entropy density in Fig.\ \ref{fig:sT4_173}  
Our results with the inclusion of Hagedorn states are able to match lattice data near the critical temperature but do not match as well at lower temperatures in Fig.\ \ref{fig:eT4} and  Fig.\ \ref{fig:sT4}.  

\begin{figure}
\centering
\includegraphics[width=3.in]{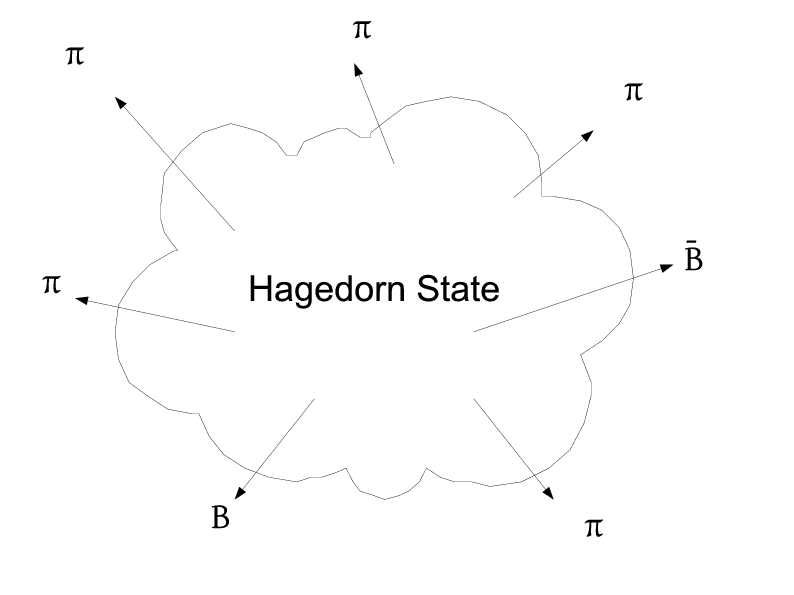}
\caption{Hagedorn states decay into multiple pions and a $X\bar{X}$ pair.   }
\label{fig:bag}
\end{figure}

Our idea is that these very massive Hagedorn states exist, as pictured in Fig.\ \ref{fig:bag}, and are so large that they decay almost immediately into multiple pions and $X\bar{X}$ pairs.  While it can be argued that Hagedorn states are more likely to 
decay
into a pair of particles: a lighter Hagedorn state and another particle, these reactions are so quick that we can consider the end results, which would be multiple particles (mostly pions).  That being said, it would be possible to put Hagedorn states into a transport approach such as UrQMD \cite{URQMD}
using binary reactions with 
possible
cross-sections as described in \cite{Pal:2005rb}.  We leave this as a challenge for the future.

Moreover, we need to consider the back reactions of multiple particles combining to form a Hagedorn state in order to preserve detailed balance.  Rate equations provide us with a perfect tool for this because there is a loss and gain term that describe both the forward and back reactions. Moreover, the state of chemical equilibrium is a fixed point of the rate equations.
The rate equations for the Hagedorn resonances $N_{i}$, pions $N_{\pi}$, and the $X\bar{X}$ pair $N_{X\bar{X}}$, respectively, are given by
\begin{eqnarray}\label{eqn:setpiHSBB}
\dot{N}_{i}&=&\Gamma_{i,\pi}\left[N_{i}^{eq}\sum_{n} B_{i,n}
\left(\frac{N_{\pi}}{N_{\pi}^{eq}}\right)^{n}-N_{i}\right]\nonumber\\
&+&\Gamma_{i,X\bar{X}}\left[ N_{i}^{eq}
\left(\frac{N_{\pi}}{N_{\pi}^{eq}}\right)^{\langle n_{i,x}\rangle} \left(\frac{N_{X\bar{X}}}{N_{X\bar{X}}^{eq}}\right)^2 -N_{i}\right]\nonumber\\
\dot{N}_{\pi }&=&\sum_{i} \Gamma_{i,\pi}  \left[N_{i}\langle n_{i}\rangle-N_{i}^{eq}\sum_{n}
B_{i, n}n\left(\frac{N_{\pi}}{N_{\pi}^{eq}}\right)^{n} \right]\nonumber\\
&+&\sum_{i} \Gamma_{i,X\bar{X}} \langle n_{i,x}\rangle\left[N_{i}-
N_{i}^{eq}
\left(\frac{N_{\pi}}{N_{\pi}^{eq}}\right)^{\langle n_{i,x}\rangle} \left(\frac{N_{X\bar{X}}}{N_{X\bar{X}}^{eq}}\right)^2\right]  \nonumber\\
\dot{N}_{X\bar{X}}&=&\sum_{i}\Gamma_{i,X\bar{X}}\left[ N_{i}- N_{i}^{eq}\left(\frac{N_{\pi}}{N_{\pi}^{eq}}\right)^{\langle n_{i,x}\rangle} \left(\frac{N_{X\bar{X}}}{N_{X\bar{X}}^{eq}}\right)^2\right].
\end{eqnarray}
The decay widths for the $i^{th}$ resonance are $\Gamma_{i,\pi}$ and $\Gamma_{i,X\bar{X}}$, the branching ratio is $B_{i,n}$ (see below), and the average number of pions that each resonance will decay into is $\langle n_{i}\rangle$.  The equilibrium values $N^{eq}$ are both temperature and chemical potential dependent.  However, here we set $\mu_b=0$.
Eq. (\ref{eqn:setpiHSBB}) can also be rewritten in terms of fugacities ($\lambda_{i}$, $\lambda_{\pi}$, and $\lambda_{X\bar{X}}$), which are found by dividing each total number by its respective equilibrium value, for example, $\lambda_{i}=\frac{N_{i}}{N_{i}^{eq}}$ (as seen for the baryon anti-baryon pairs in \cite{Noronha-Hostler:2007fg}).  
Additionally, a discrete spectrum of Hagedorn states is considered, which is separated into mass bins of 100 MeV.  Each bin is described by its own rate equation.

The branching ratios, $B_{i,n}$, are the probability that the $i^{th}$ Hagedorn state will decay into $n$ pions.  Since we are dealing with probabilities, then $\sum_{n}B_{i,n}=1$ must always hold. 
In order to include a distribution for our branching ratios we assume that they follow a Gaussian distribution for the reaction $HS\leftrightarrow n\pi$ 
\begin{equation}
B_{i, n}\approx
\frac{1}{\sigma_{i}\sqrt{2\pi}}e^{-\frac{(n-\langle n_{i}\rangle)^{2}}{2\sigma_{i} ^{2}}},
\end{equation}
which has its peak centered at $\langle n_{i}\rangle$ and the width of the distribution is $\sigma^2$.  Assuming a statistical, micro-canonical branching for the decay of Hagedorn states, we can take a linear fit to the average number of pions in Fig.\ 1 in
Ref.\ \cite{Greiner:2004vm} (multiplying $\pi^+$ by three to include all pions) to find $\langle n_{\pi}\rangle$ such that $\langle n_i\rangle=0.9+1.2\frac{m_{i}}{m_{p}}$ is the average pion number that each
Hagedorn state decays into.  Within the microcanonical model a Hagedorn state is defined by its mass and corresponding volume where the volume is taken as $V=m_{i}/\varepsilon$. The mean energy density of a Hagedorn state is $\varepsilon$ (taken as $\varepsilon=0.5\frac{GeV}{fm^3}$). Further discussions regarding this can be found in \cite{Greiner:2004vm,Liu}.
The width of the distribution is $\sigma^{2}_{i}=(0.5\frac{m_{i}}{m_{p}})^{2}$.  Both our choice in $\langle n_i\rangle$ and $\sigma^{2}_i$ roughly match the canonical description in \cite{Becattini:2004rq}. 

Furthermore, we have the condition that each Hagedorn resonance must decay into at least 2 pions.  Because of the nature of a Gaussian distribution there is a non-zero probability that a Hagedorn state can decay into less than 2 pions.  Therefore, we calculate the percentage of the distribution that falls below 2 pions and redistribute that over $n\geq 2$ so that $\sum_{n}B_{i,n}=1$.  This in turn leads to a new $\langle n_{i}\rangle$ and  $\sigma^2_i$, which we find by calculating $\langle n_{i}\rangle=\sum_{n}nB_{i,n}$ and $\sigma^2_i=\langle n^2_i\rangle-\langle n_i\rangle ^2$. Thus, we after normalize for the cutoff $n\geq 2$, we have $\langle n_{i}\rangle\approx 3 - 34$ and $\sigma_{i}^2\approx0.8 - 510$.  

For the average number of pions when a $X\bar{X}$ pair is present, we again refer to the micro-canonical model in \cite{Greiner:2004vm,Liu}.  We use $\langle n_{\pi}\rangle$ but then readjust it to the average pion number according to Fig.\ 2 in
Ref.\ \cite{Greiner:2004vm} for when a baryon anti-baryon pair is present (there the distribution is for a resonance of mass $m=4$ GeV).  Thus, 
\begin{equation}\label{eqn:nfit}
    \langle n_{i,x}\rangle=\left(\frac{2.7}{1.9}\right)\left(0.3+0.4m_i\right)\approx 2-7.
\end{equation}
where $m_i$ is in GeV. In this paper we do not consider a distribution but rather only the average number of pions when a $X\bar{X}$ pair is present.  We  assume that $\langle n_{i,x}\rangle=\langle n_{i,p}\rangle=\langle n_{i,k}\rangle=\langle n_{i,\Lambda}\rangle=\langle n_{i,\Omega}\rangle$ for when a kaon anti-kaon pair, $\Lambda\bar{\Lambda}$, or  $\Omega\bar{\Omega}$ pair is present.  Ideally, $\langle n_{i,k}\rangle$, $\langle n_{i,\Lambda}\rangle$, and $\langle n_{i,\Omega}\rangle$ should be derived separately and will be done in a future paper using a canonical model \cite{ref:max}.

We used a linear fit for the total decay width similar to that used in
Ref.\ \cite{Senda}. The total decay width 
\begin{equation}
\label{HSdecaywidth}
\Gamma_{i}=0.15m_{i}-0.0584
\end{equation}
($\Gamma_i $ and $m_i$ in terms of GeV), which ranges from $\Gamma_{i}=250\;\mathrm{MeV\;to}\;1800$ MeV, is a linear fit extrapolated from
the data in Ref.\ \cite{Eidelman:2004wy}. However, in Eq.\ (\ref{eqn:setpiHSBB}) the total decay width is separated into two parts: one for the reactions $HS\leftrightarrow n\pi$, $\Gamma_{i,\pi}$, and one for the reaction in Eq.\ (\ref{eqn:decay}), $\Gamma_{i,X\bar{X}}$, whereby 
$\Gamma_{i}=\Gamma_{i,\pi}+\Gamma_{i,X\bar{X}}$.  Then relative decay width $\Gamma_{i,X\bar{X}}$ is the average number of $X\bar{X}$ in the system $\langle X\rangle$ multiplied by the total decay width $\Gamma_{i}$.  Essentially, a fraction of the decay of the $i^{th}$ Hagedorn state goes into $X\bar{X}$ (set by the number of $X\bar{X}$ the $i^{th}$ Hagedorn state on average decays into) and the remainder goes into pions.  

\begin{figure}[b]
\centering
\includegraphics[width=3.in]{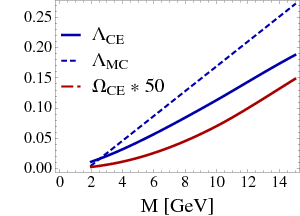}
\caption{Average number of $\Lambda$'s and $\Omega$'s.  The $\Omega$'s are calculated within our canonical ensemble and the $\Lambda$'s are calculated both in our canonical ensemble and a micro-canonical ensemble.} \label{fig:lamom}
\end{figure}

We find $\langle p\rangle$  by linearly fitting the proton in Fig.\ 2 in Ref.\ \cite{Greiner:2004vm} so that
\begin{eqnarray}\label{eqn:gamfit}
  p &=& 0.058\;m_{i}-0.10 
%  neutrons &=& 0.048\;m_{i}-0.052\;.
\end{eqnarray}
where $m_i$ is in GeV and  $\langle p\rangle\approx 0.01 - 0.6$.  Thus, $\Gamma_{i,p\bar{p}}$ is between $3$ and $1000$ MeV. Clearly,
$\Gamma_{i,\pi}$ is then
$\Gamma_{i,\pi}=\Gamma_{i}-\langle p\rangle\Gamma_{i,\pi}$.
Analogously for the kaons, the decay width is $\Gamma_{i,K\bar{K}}=\langle K\rangle \Gamma_{i}$ where 
\begin{equation}
K^{+}=0.075\;m_{i}+0.047
\end{equation}
where $m_i$ is in GeV,
which is also taken from Fig.\ 2 in Ref.\ \cite{Greiner:2004vm}. We find that
$\langle K\rangle=0.2$ to $0.95$ \cite{Liu,Greiner:2004vm}. Thus, $\Gamma_{i,K\bar{K}}$ is between $50$ and $1700$ MeV.

For $\Lambda$ we use a canonical model assuming that the baryon number $B=0$, the strangeness $S=0$, and the electrical charge $Q=0$ in order to calculate the average lambda number.  The results of this are shown in Fig.\ \ref{fig:lamom}.  We find that our $\langle\Lambda\rangle$ is lower than that from the micro-canonical ensemble in \cite{Greiner:2004vm}, which is also shown in  Fig.\ \ref{fig:lamom}.  This corresponds to a decay width of $\Gamma_{i,\Lambda\bar{\Lambda}}=3-250$ MeV.

Furthermore, the average number of $\Omega$'s is also shown in Fig.\ \ref{fig:lamom} from our canonical model again assuming that the baryon number $B=0$, the strangeness $S=0$, and the electrical charge $Q=0$.  In Fig.\ \ref{fig:lamom} we multiple  $\langle\Omega\rangle$ in order to better view the results. The resulting decay width is $\Gamma_{i,\Omega\bar{\Omega}}=.01-4$ MeV.

The equilibrium values are found using a statistical model \cite{StatModel}, which includes 104 particles from the the PDG \cite{Eidelman:2004wy} (only light and strange particles). As in \cite{StatModel}, we also consider the effects of feeding (the contributions of higher lying resonances such as the $\rho$ or $\omega$ resonances on the number of ``pions'' in our system, i.e., $N_{\pi}^{eq}$ includes ``all" the pions from resonances from the PDG \cite{Eidelman:2004wy}).  Feeding is also considered for the protons, kaons, and lambdas. Additionally, throughout this paper our initial conditions are the various fugacities at $t_0$ (at the point of the phase transition into the hadron gas phase)

%
%CG: how does alpha and beta really enter into the following eq. 11 ???
% what then means s_Had+Hs(T_0) ??? please explain !
% 

\begin{equation}
\label{initcond}
\alpha\equiv\lambda_{\pi}(t_0) \, , \, \beta_{i}\equiv\lambda_{i}(t_0) \, , \mbox{and}\,   \phi\equiv\lambda_{X\bar{X}}(t_0) \, \, , 
\end{equation}
which are chosen by holding the contribution to the total entropy from the Hagedorn states and pions constant i.e. 
\begin{eqnarray}\label{entrcont}
s_{Had}(T_{0},\alpha)V(t_{0})+s_{HS}(T_{0},\beta_{i})V(t_{0})\nonumber\\
=s_{Had+HS}(T_{0})V(t_{0})=const
\, \, .
\end{eqnarray} 
and the corresponding initial condition configurations we choose later can be seen in Tab.\ \ref{tab:IC}. $s_{Had}(T_{0},\alpha)$ is the entropy density at the initial temperature, i.e., the critical temperature multiplied by our choice in $\alpha$.  Because the hadron resonance is dominated by pions we can assume that $\alpha$ represents the initial fraction of pions in equilibrium.  $s_{HS}(T_{0},\beta_{i})$ represents the entropy contribution from the Hagedorn states at $T_c$ multiplied by the initial fraction of Hagedorn states in equilibrium.  We hold $\alpha$ constantly and then find the appropriate $\beta_i$.
The volume expansion, $V(t)$ is discussed in detail following section entitled 'Expanding Fireball'.

\section{Chemical Equilibration Time Estimate}\label{tau}

As a starting point of our analysis, 
we first estimate the chemical equilibration time of the $X\bar{X}$ by looking at the fugacity of the $X\bar{X}$ rate equation, i.e., Eq.\ (\ref{eqn:setpiHSBB}) can be rewritten in terms of $\lambda$ as shown for $B\bar{B}$ in Eq. (3) in \cite{Noronha-Hostler:2007fg}, when both the pions and Hagedorn states are held constant.  
\begin{figure}[h]
\begin{center}
\includegraphics[width=3.in]{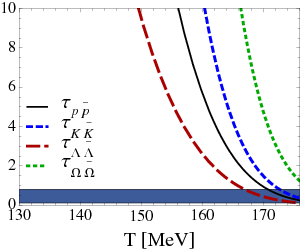}\\
\includegraphics[width=3.in]{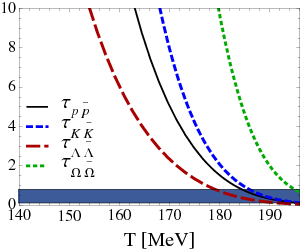}
\end{center}
\caption{Comparison of the chemical equilibrium times for protons, kaons, and lambdas when $\alpha=1$ and $\beta_{i}=1$ where $T_H=176$ MeV (top) and $T_H=196$ MeV (top). The gray band is the range of chemical equilibrium times for the Hagedorn states (see Tab.\ \ref{tab:tau}). }
\label{fig:taubb}
\end{figure}
The $X\bar{X}$ rate equation then becomes
\begin{equation}\label{eqn:lambbfirst}
\dot{\lambda}_{X\bar{X}}=\sum_{i}\Gamma_{i,X\bar{X}}\frac{N_{i}^{eq}}{N_{X\bar{X}}^{eq}}\left( \beta_{i}- \alpha^{\langle n_{i,x}\rangle} \lambda_{X\bar{X}}^2\right), 
\end{equation}
which we can integrate   
\begin{equation}\label{eqn:lambba1}
\lambda_{X\bar{X}}=\zeta\left[\frac{\left(\frac{\phi+\zeta}{\phi-\zeta}\right) e^{\frac{2t}{\tau_{X\bar{X}}}}+1}{\left(\frac{\phi+\zeta}{\phi-\zeta}\right) e^{\frac{2t}{\tau_{X\bar{X}}}}-1}\right]
\end{equation}
where 
\begin{equation}\label{eqn:taubb}
\tau_{X\bar{X}}\equiv\frac{N_{X\bar{X}}^{eq}}{\sqrt{\sum_{i}\Gamma_{i,X\bar{X}} N_{i}^{eq}\beta_{i}}\sqrt{\sum_{i}\Gamma_{i,X\bar{X}} N_{i}^{eq}\alpha^{\langle n_{i,x}\rangle}}}\;,
\end{equation}
$\zeta\equiv\sqrt{\frac{\sum_{i}\Gamma_{i,X\bar{X}} N_{i}^{eq}\beta_{i}}{\sum_{i}\Gamma_{i,X\bar{X}} N_{i}^{eq}\alpha^{\langle n_{i,x}\rangle}}}$, and $\lambda_{X\bar{X}}(0)\equiv\phi$.  Substituting in $\alpha=1$ and $\beta_{i}=1$ when the pions and Hagedorn states are in chemical equilibrium, we rederive Eq.\ (7) in Ref.\ \cite{Noronha-Hostler:2007fg}
\begin{equation}\label{eqn:ateqbb}
\tau_{X\bar{X}}=\frac{N_{X\bar{X}}^{eq}}{\sum_{i}\Gamma_{i,X\bar{X}}N_{i}^{eq}},
\end{equation}
which is shown in Fig.\ \ref{fig:taubb}. From Eq.\ (\ref{eqn:ateqbb}) we see that the time scale has an indirect dependence on the decay width.  Since the decay width has a linear dependence on the mass, the time scale decreases when more Hagedorn states are included.  However, $N_i^{eq}$ also decreases with increasing mass so above a certain point very many Hagedorn states need to be included in order to see an effect in the time scale.  Furthermore, the chemical equilibrium values have a dependence on the temperature, which makes the time scale shortest for the highest temperatures. 

\begin{figure*}
\begin{minipage}{0.45\linewidth}
\centering
\includegraphics[width=3.in]{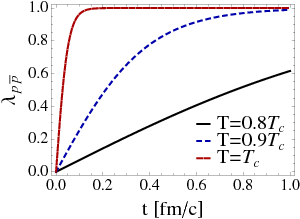} \\
\includegraphics[width=3.in]{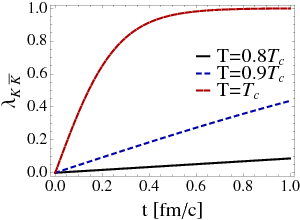} \\
\includegraphics[width=3.in]{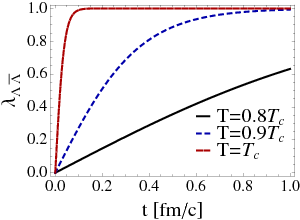} 
\caption{Graph of the number of proton anti-proton pairs, kaon anti-kaon pairs, and lambda anti-lambda pairs when both
the resonances and pions are held in equilibrium for $T_H=176$ MeV.}
\label{fig:HSpireineq176} 
\end{minipage}
\hspace{0.5cm}
\begin{minipage}{0.45\linewidth}
\includegraphics[width=3.in]{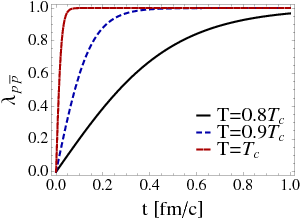} \\
\includegraphics[width=3.in]{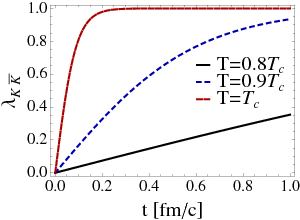} \\
\includegraphics[width=3.in]{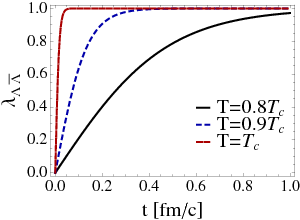} 
\caption{Graph of the number of proton anti-proton pairs, kaon anti-kaon pairs, and lambda anti-lambda when both
the resonances and pions are held in equilibrium for $T_H=196$ MeV.}
\label{fig:HSpireineq196} 
\end{minipage}
\end{figure*}

In Fig.\ \ref{fig:HSpireineq176} and Fig.\ \ref{fig:HSpireineq196} we hold the Hagedorn states and pions and let the $X\bar{X}$ pairs reach chemical equilibrium.  That means that in Eq.\ (\ref{eqn:setpiHSBB}) we set $N_{\pi}=N_{\pi}^{eq}$ and $N_{i}=N_{i}^{eq}$ in the $\dot{N}_{X\bar{X}}$ equation.  Fig.\ \ref{fig:HSpireineq176} shows the results for $p\bar{p}$, $K\bar{K}$, and $\Lambda\bar{\Lambda}$, respectively, for $T_H=176$ MeV and  Fig.\ \ref{fig:HSpireineq196} shows the same for $T_H=196$ MeV.  In all cases the temperature is held constant while the rate equations are solved over time.  At $T=T_c$ all $X\bar{X}$ reach chemical equilibrium almost immediately (on the order of $t<0.2fm/c$).  As T is decreased the chemical equilibrium time obviously increases, which is clear from Fig.\ \ref{fig:taubb}.

Even as the temperature is lowered we still see quick chemical equilibrium times.  For the $p\bar{p}$ and $\Lambda\bar{\Lambda}$ pairs at $T=0.9T_c$ the chemical equilibrium time is still about $t<1 fm/c$.  The $K\bar{K}$ pairs do have a slower chemical equilibrium time due to their larger chemical equilibrium abundances, which is directly related to the chemical equilibration time through Eq.\ (\ref{eqn:setpiHSBB}).  
This again represents the main idea, which is the importance of potential
Hagedorn states in understanding fast chemical equilibration of hadrons close and below $T_c$. The Hagedorn states increase dramatically in number close to
the critical temperature and, thus, by its subsequent decay and re-population they
will quickly produce the various hadronic particles.

The equilibration of $X\bar{X}$ pairs then shown in Fig.\ \ref{fig:HSpireineq176} and Fig.\ \ref{fig:HSpireineq196} where the analytical result in Eq.\ (\ref{eqn:lambba1}) matches the numerical result exactly. From  Fig.\ \ref{fig:HSpireineq176} and Fig.\ \ref{fig:HSpireineq196} it can be seen that all $X\bar{X}$ pairs equilibrate quickly close to the critical temperature $\tau<1\frac{fm}{c}$. Clearly, though, as the temperature decreases the chemical equilibration time lengthens.  However, at $T_H=196$ MeV chemical equilibrium is still reached quickly, $\tau<1\frac{fm}{c}$.

\section{Analytical Results: Pions and Hagedorn Resonances}\label{ar}

While the chemical equilibration time derived in the previous section is a good estimate, it can only be strictly applied when the pions and Hagedorn states are assumed to stay in chemical equilibrium at a constant temperature (Fig.\ \ref{fig:HSpireineq176} and Fig.\ \ref{fig:HSpireineq196}).  Otherwise, non-linear effects that appear when the pions and Hagedorn states are allowed to equilibrate appear.  

\begin{figure}
\begin{center}
\includegraphics[width=3.in]{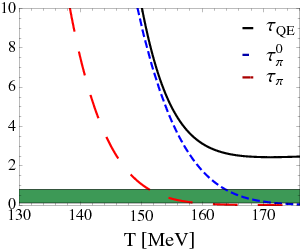}\\
\includegraphics[width=3.in]{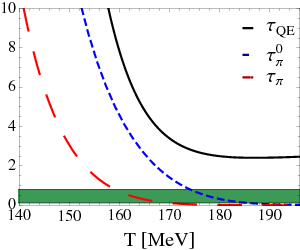} 
\end{center}
\caption{Comparison of the chemical equilibration times of the pions to the total chemical equilibration time for $T_H=176$ MeV (top) and $T_H=196$ MeV (bottom).}
\label{fig:taupi}
\end{figure}

To understand the dynamics in more detail, we consider the simplified case when the Hagedorn resonances decay only into pions $HS\leftrightarrow n\pi$, which gives
\begin{eqnarray} \label{eqn:setpiHS}
\dot{N}_{i}&=&\Gamma_{i} \left[N_{i}^{eq} \sum_{n=2}B_{i,n}
\left(\frac{N_{\pi }}{N_{\pi}^{eq}}\right)^{n}-N_{i}\right]  \nonumber\\
\dot{N}_{\pi}&=&\sum _{i} \Gamma_{i} \left[N_{i}\langle n_i\rangle-
 N_{i}^{eq} \sum _{n=2}  
B_{i,n}n\left(\frac{N_{\pi }}{N_{\pi}^{eq}}\right)^{n} \right].
\end{eqnarray}
Assuming that the pions and the Hagedorn states described in Eq.\ (\ref{eqn:setpiHS}) are then allowed to equilibrate near $T_{c}$ in a static system, we are able to derive analytical solutions, the derivation of which is shown in detail in Appendix \ref{app}.  For the analytical solutions we divide the chemical equilibration into three stages, the chemical equilibration times of which are shown in Tab.\ \ref{tab:tau}. The first stage (described by $\tau_\pi^0$ in Tab.\ \ref{tab:tau}) of the evolution is dominated by the chemical equilibration of the pions when the pions are still far away from their chemical equilibrium values.  After the pions are close to chemical equilibrium, new dynamics take over, which are described by $\tau_\pi$ in Tab.\ \ref{tab:tau} and Fig.\ \ref{fig:taupi}. 

\begin{table}
\begin{center}
 \begin{tabular}{|c|c|cc|}
 \hline
   &  & $M_{2 GeV}$  & $M_{12 GeV}$\\
  HS & $\tau_{i}=1/\Gamma_{i}$ & $0.8\frac{\mathrm{fm}}{c}$ & $0.1\;\frac{\mathrm{fm}}{c}$\\
 \hline
    &  &  & \\
    &  &  $0.95 T^{BMW}_c$& $0.95 T^{RBC}_c$ \\
 \hline
 $\lambda_{\pi}\approx 0$ & $\tau_{\pi}^{0}\equiv\frac{N_{\pi}^{eq}}{\sum_{i} \Gamma_{i} N^{eq}_{i} \langle n_{i}\rangle \beta_{i}}$ & $0.5\frac{\mathrm{fm}}{c}$ & $0.1\frac{\mathrm{fm}}{c}$\\
 $\lambda_{\pi}\approx 1$ & $\tau_{\pi}\equiv\frac{N_{\pi}^{eq}}{\sum_{i} \Gamma_{i} N^{eq}_{i} \langle n_{i}^2\rangle}$ & $0.01\frac{\mathrm{fm}}{c}$ & $0.003\frac{\mathrm{fm}}{c}$\\
 QE & $\tau^{QE}_{\pi}\equiv\frac{N_{\pi}^{eq}}{\sum_{i} \Gamma_{i} N^{eq}_{i} \sigma_{i}^2}+\frac{\sum_{QE}N_{i}^{eq}\langle n_{i}^2\rangle}{\sum_{i}\Gamma_{i}N_{i}^{eq}\sigma_{i}^2}$ & $1.7\frac{\mathrm{fm}}{c}$ & $1.6\frac{\mathrm{fm}}{c}$\\
 \hline 
 \end{tabular}
 \end{center}
 \caption{Chemical equilibration times from analytical estimates where QE is quasi-equilibrium at $95\%$ of each respective $T_H$.}\label{tab:tau}
 \end{table}

In both Stage 1 and 2 the equilibration of the Hagedorn states is set by the dynamics of the pions.  Finally, in Stage 3 the pions, which are already almost in chemical equilibrium, reach a quasi-equilibrium state with the Hagedorn states.  Quasi-equilibrium is reached when at least one species of Hagedorn states has succeeded it's chemical equilibrium time scale determined from the inverse of its decay width, i.e., $\tau_i=1/\Gamma_i$. Since the heaviest Hagedorn states have the shortest $\tau_i$'s, then quasi-equilibrium is reached when $\tau_i$ of the heaviest Hagedorn state is surpassed.  During this stage non-linear affects take over and, thus, a longer time scale, $\tau_\pi^{QE}$, is seen.  While this time scale may appear long, both the pions and Hagedorn states are so close to chemical equilibrium that they are within roughly $10\%$ (depending on the initial conditions) or less of their chemical equilibrium values before quasi-equilibrium is even reached.  The detailed calculations are shown in the Appendix. 

Therefore, the most important chemical equilibration time is then that from the pions in Stage 1, i.e., $\tau_\pi^0$.  The time scale from Stage 2 is so short that it is not of much importance.  Additionally, by the time that Stage 3 is reached both the pions and Hagedorn states are essentially in chemical equilibrium and, therefore, the non-linear affects do not play a large role in the overall chemical equilibration time.   One can see this more clearly in the top panel of  Fig. \ref{fig:pifree} in Appendix \ref{app} where the pions and heavier Hagedorn states are extremely close to chemical equilibrium, while the lighter Hagedorn states are still only moderately 
close to their chemical equilibrium values. Therefore, the Hagedorn states and pions are able to be roughly in chemical equilibrium on the order of $<1\frac{fm}{c}$ according to our analytical solution when held at a constant temperature. 

This also applies to the $K\bar{K}$ reaction $HS\leftrightarrow n\pi+K\bar{K}$ as shown in the bottom panel of  Fig. \ref{fig:pifree} in Appendix \ref{app}.  The time scale for the pions and Hagedorn states are slightly longer when the $K\bar{K}$ pairs are present.  The same goes for the estimated chemical equilibration time of the $K\bar{K}$ pairs in the previous section, $\tau_{K\bar{K}}$.

\section{Expanding Fireball}\label{expansion}

In order to include the cooling of the fireball we need to find a relationship between the temperature and the time, i.e., $T(t)$.  To do this we apply a Bjorken expansion for which the total entropy is held constant
\begin{equation}\label{eqn:constrain}
\mathrm{const.}=s(T)V(t)\sim\frac{S_{\pi}}{N_{\pi}}\int \frac{dN_{\pi}}{dy} dy.
\end{equation}
where $s(T)$ is the entropy density of the hadron gas with volume corrections.

\begin{figure}
\centering
\includegraphics[width=3.in]{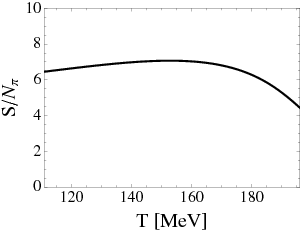} 
\caption{Entropy per pion for a hadron gas in chemical equilibrium within the fireball ansatz.} \label{fig:sn}
\end{figure}

The total number of pions in the $5\%$ most central collisions, $\frac{dN_{\pi}}{dy}$, can be found from experimental
results in \cite{Bearden:2004yx}.  There they
found the phase-space yields for the pions $\pi^{+}$
($292.0$) and $\pi^{-}$ ($290.9$) using a
Gaussian fit for yields as a function of the rapidity
$\frac{dN_{\pi}}{dy}$ where we used the rapidity range $y=\pm 0.5 $.  We then assumed that the number of
$\pi^{0}$'s were also in that same range and took the average of the
two to find $291.5$.  Thus, our total pion number is
$\sum_{i}N_{\pi^{i}}=\int_{-0.5}^{0.5} \frac{dN_{\pi}}{dy} dy=874$.
While for a gas of non-interacting Bose gas of massless pions $S_{\pi}/N_{\pi}=3.6$, we do have a mass for a our pions, so we must adjust $S_{\pi}/N_{\pi}$ accordingly.  In \cite{Greiner:1993jn} it was shown that when the pions have a mass the ratio changes and, therefore, the entropy per pion is close to $S_{\pi}/N_{\pi}\approx5.5$. The actual $S_{\pi}/N_{\pi}$ in our model is shown in Fig.\ \ref{fig:sn} where $S_{\pi}/N_{\pi}\approx 6$, which is only slightly higher.

\begin{figure}[h]
\centering
\includegraphics[width=3.in]{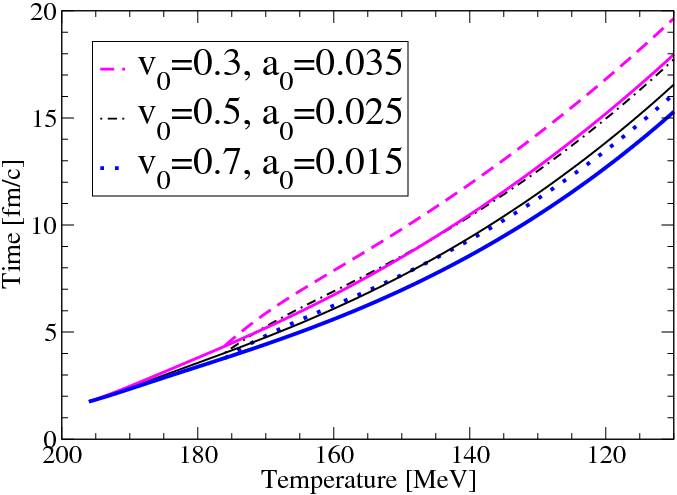}
\caption{The temperature-time relationship is directly linked to the average transversel velocity chosen in Eq.\ (\ref{eqn:bjorken}) within the fireball model ansatz.}
\label{fig:temptime}
\end{figure}

\begin{figure}[ht]
\centering
\includegraphics[width=3.in]{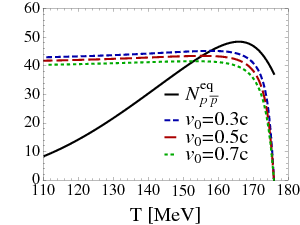} \\
\includegraphics[width=3.in]{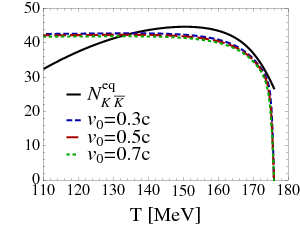} \\
\includegraphics[width=3.in]{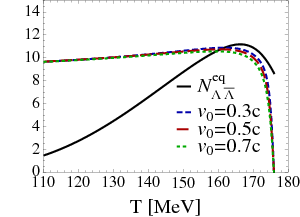} 
\caption{Results for the $p\bar{p}$, $K\bar{K}$, and $\Lambda\bar{\Lambda}$ when the pions and Hagedorn resonances are held in equilibrium for $T_H=176$ MeV.} \label{fig:Ex.BBrepiineq176}
\end{figure}

\begin{figure}[ht]
\centering
\includegraphics[width=3.in]{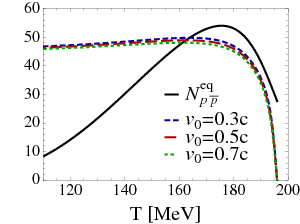} \\
\includegraphics[width=3.in]{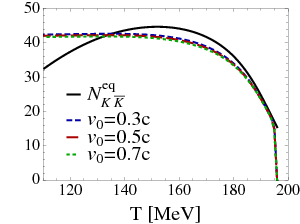}  \\
\includegraphics[width=3.in]{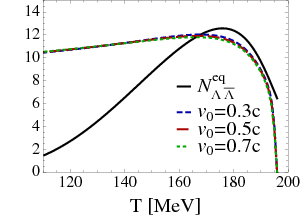} 
\caption{Results for the $p\bar{p}$, $K\bar{K}$, and $\Lambda\bar{\Lambda}$ when the pions and Hagedorn resonances are held in equilibrium for $T_H=196$ MeV.} \label{fig:Ex.BBrepiineq196}
\end{figure}

The effective volume at mid-rapidity can be parametrized as a function of time.  We do this by using a Bjorken expansion and including accelerating radial flow.  
The volume term is then
\begin{equation}\label{eqn:bjorken}
V(t)=\pi\;ct\left(r_{0}+v_{0}(t-t_{0})+\frac{1}{2}a_{0}(t-t_{0})^2 \right)^2
\end{equation}
where the initial radius is $r_{0}(t_0)=7.1$ fm for $T_H=196$ and the corresponding $t_{0}^{(196)}\approx2 fm/c$. For $T_H=176$ we allow for a longer expansion before the hadron gas phase is reached and, thus, calculate the appropriate $t_0^{(176)}$ from the expansion starting at $T_H=196$, which is $t_0^{(176)}\approx 4 fm/c$ (there is a slightly variation dependent on the choice of $v_{0}$ and $a_{0}$). The $T(t)$ relation is shown in Fig.\ \ref{fig:temptime}, which has almost no effect on the results as seen later on in Fig.\ \ref{fig:Ex.BBrepiineq176} and Fig.\ \ref{fig:Ex.BBrepiineq196}. Therefore, we choose $v_0=0.5 $ and $a_0=0.025 $ for the remainder of this paper. 
The relation depicted
allows to translate the later shown figures labeled by the effective 
global temperature of the evolving system directly into the evolving
system time.

\begin{figure}[h]
\centering
\includegraphics[width=3.in]{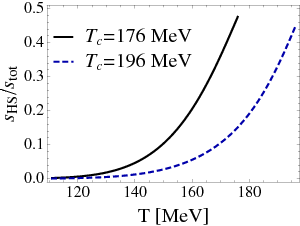}
\caption{Ratio of the entropy of the Hagedorn states to the total entropy.}
\label{fig:entropycomp}
\end{figure}

\begin{figure}[h]
\centering
\includegraphics[width=3.in]{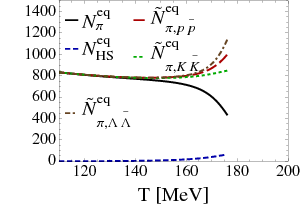}\\
\includegraphics[width=3.in]{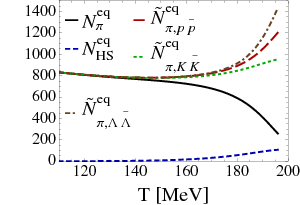}
\caption{Comparison of the effective pion numbers when $T_H=176$ MeV (top) or  $T_H=196$ MeV (bottom).} \label{fig:effnum}
\end{figure}

\begin{figure}[h]
\centering
\begin{tabular}{c}
\includegraphics[width=3.in]{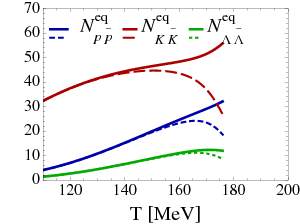}\\
\includegraphics[width=3.in]{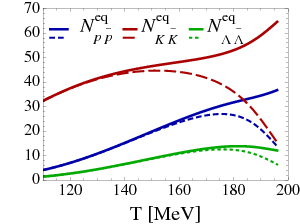}
\end{tabular}
\caption{Comparison of the total number of $X\bar{X}$ and their effective numbers when $T_H=176$ MeV (top) or  $T_H=196$ MeV (bottom).} \label{fig:density}
\end{figure}

Because the volume expansion depends on the entropy according to Eq.\ (\ref{eqn:constrain}) and the Hagedorn resonances contribute strongly to the entropy only close to the critical temperature (see Fig.\ \ref{fig:entropycomp}), the equilibrium values actually decrease with increasing temperature close to $T_{c}$ for the hadrons as seen in Fig.\ \ref{fig:effnum} and Fig.\ \ref{fig:density}.  One can clearly see from Fig.\ \ref{fig:entropycomp} that the Hagedorn states contribute strongly close to $T_c$ down to about $80\%$ of $T_c$.

Therefore, one has to include the 
potential
contribution of the Hagedorn resonances to the pions
as in the case of standard hadronic resonances,
e.g. a $ \rho $-meson decays dominantly into
two pions and, thus, accounts for them by a factor two.  
This is similar to what was done in Appendix \ref{app} in Eq.\ (\ref{eqn:effpions}).   Including the Hagedorn state contribution, we arrive at our effective number of pions
\begin{eqnarray}\label{eqn:effpi}
\tilde{N}_{\pi,X\bar{X}}&=&N_{\pi}\nonumber\\
&+&\sum_{i}N_{i}\left[\left(1-\langle X_i\rangle\right)\langle n_{i}\rangle +\langle X_i\rangle\langle n_{i,x}\rangle\right]
\end{eqnarray}
which are shown in Fig.\ \ref{fig:effnum}. In  Fig.\ \ref{fig:effnum} we see that after the inclusion of the effective pion numbers that the number of pions only decreases with decreasing temperature.  Furthermore, in Fig.\ \ref{fig:effnum} the total number of Hagedorn states, $\sum_{i}N_i^{eq}$ is also shown.  While there are 
by far
fewer Hagedorn states present than pions, we see that they are important because of their large contribution to the entropy density as shown in Fig.\ \ref{fig:entropycomp}.
The reason that the effective number of pions increase close to $T_c$ is due to the large number of pions that the heavy Hagedorn states decay into.  If $\langle n_i\rangle$ was smaller or no longer linear than it could be possible that the effective number of pions would remain constant.

Moreover, it is useful to consider the effective number of $X\bar{X}$ pairs
\begin{eqnarray}\label{eqn:effbbkk}
\tilde{N}_{X\bar{X}}&=&N_{X\bar{X}}+\sum_{i}N_{i}\langle X_i\rangle
\end{eqnarray}
because Hagedorn states also contribute strongly to the $X\bar{X}$ pairs close to $T_c$ as seen in Fig. \ref{fig:density}. Again we see that only the effective number of $X\bar{X}$ pairs have consistent decreasing behaviour with decreasing temperature whereas without the Hagedorn state contributions we see a decrease close to $T_c$.

Along with the expansion we also must solve these rate equations, Eq. (\ref{eqn:setpiHSBB}), numerically \footnote{We solve our coupled non-linear differential equations using NDSolve in Mathematica}.  We start with various initial conditions, as mentioned previously, that are described by $\alpha$, $\beta_i$, and $\phi$ (see table II).  The initial temperature is the respective critical temperature and we end the expansion at $T=110$ MeV, a global kinetic freezeout temperature.

For the remainder of this paper we include only results for an expanding fireball, which are solved numerically.  As an initial test we hold both the pions and Hagedorn states in chemical equilibrium and allow just $X\bar{X}$ to equilibrate as seen in Fig.\ \ref{fig:Ex.BBrepiineq176} and Fig.\ \ref{fig:Ex.BBrepiineq196}.   The black solid line in each graph is the chemical equilibrium abundances and the colored lines are the dynamical calculations for various expansions that follow the $T(t)$ shown in Fig.\ \ref{fig:temptime}.
We see that regardless of our volume expansion they all quickly approach equilibrium.  
In Fig.\ \ref{fig:Ex.BBrepiineq176} and Fig.\ \ref{fig:Ex.BBrepiineq196} the $X\bar{X}$ all reach chemical equilibrium almost immediately, well before  $0.9 T_c$ the chemnical equilibration time is $<1\frac{fm}{c}$.  The only exception is the $K\bar{K}$ pairs for $T_H=176$ MeV.  However, we see later on that the $K/\pi$ ratio matches the data.  

\begin{table}
\begin{center}
 \begin{tabular}{|c|c|c|c|}
 \hline
 & & & \\
   & $\alpha=\lambda_{\pi}(t_0)$ & $\beta_{i}=\lambda_i(t_0)$ & $\phi=\lambda_{X\bar{X}}(t_0)$ \\
    & & & \\
 \hline
$IC_1$ & 1 & 1 & 0 \\
$IC_2$ & 1 & 1 & 0.5 \\
$IC_3$ & 1.1 & 0.5 & 0 \\
$IC_4$ & 0.95 & 1.2 & 0 \\
 \hline
 \end{tabular}
 \end{center}
 \caption{Initial condition configurations, recalling Eq.\ (\ref{initcond})}\label{tab:IC}
 \end{table}

\begin{figure*}
\begin{minipage}{0.45\linewidth}
\centering
\includegraphics[width=3.in]{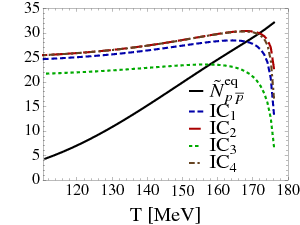} \\
\includegraphics[width=3.in]{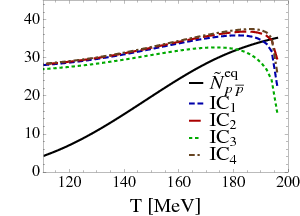} 
\caption{Results for the p's and pions with various initial conditions for $T_H=176$ MeV.} \label{fig:pp176}
\end{minipage}
\hspace{0.5cm}
\begin{minipage}{0.45\linewidth}
\includegraphics[width=3.in]{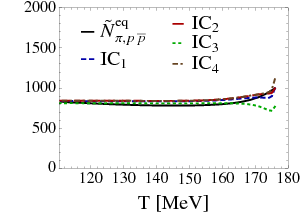}  \\
\includegraphics[width=3.in]{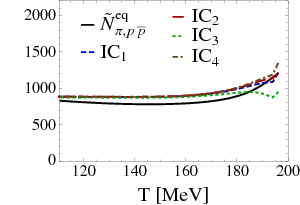} 
\caption{Results for the p's and pions with various initial conditions for $T_H=196$ MeV..} \label{fig:pp196}
\end{minipage}
\end{figure*}

\begin{figure}
\centering
\includegraphics[width=3.in]{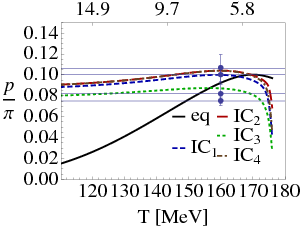}  \\
\includegraphics[width=3.in]{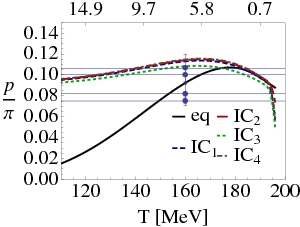} 
\caption{Results for the ratio of p's with various initial conditions. Note that for STAR $p\pi^-0.11$ and  $\bar{p}\pi^-=0.082$. Along the top axis of each graph the corresponding time is shown in $fm/c$.} \label{fig:pppi}
\end{figure}

More interestingly, we consider the case when the pions, Hagedorn states, and $X\bar{X}$ all are allowed to chemical equilibrate.  We then vary the initial conditions and observe their effects.  The results for $p\bar{p}$ pairs are shown in 
Fig.\ \ref{fig:pp176} and Fig.\ \ref{fig:pp196}.  In Fig.\ \ref{fig:pp176} and Fig.\ \ref{fig:pp196} we show the evolution of both the $p\bar{p}$ pairs and the pions for the reaction $n\pi\leftrightarrow HS\leftrightarrow n\pi+X\bar{X}$. Note that in all the following figures the effective numbers are shown so that the contribution of the Hagedorn states is included.

One can see that the chemical equilibration time does depend slightly on our choice of $\beta_{i}$, i.e., a larger $\beta_{i}$ means a quicker chemical equilibration time.   For instance, if the Hagedorn states were overpopulated coming out of the QGP phase than chemical equilibrium times would be slightly shorter.  
However, even when the Hagedorn resonances start underpopulated the $p\bar{p}$ pairs are able to reach chemical equilibrium immediately.  Additionally, when the $p\bar{p}$ pairs start at about half their chemical equilibrium values, it only helps the $p\bar{p}$ pairs to reach equilibrium at a slightly higher temperature (on the order of a couple of MeV).  Additionally, we see a greater dependence on $\beta_{i}$ for $T_H=176$ MeV than for $T_H=176$ MeV. 
Throughout the evolution we see from the pions that they remain roughly in chemical equilibrium.  Thus, our initial analytical approximation appears reasonable.  

In Fig.\ \ref{fig:pppi} the ratio of protons's to $\pi$'s is shown.  We also compare our results to that of experimental data.  We see that for $T_H=176$ MeV that our results enter the band of experimental data before $T=170$ MeV and remain there throughout the entire expansion regardless of the initial conditions. However, for  $T_H=176$ MeV the results are slightly different.  In this case, the ratios match the experimental data early on at around $T=190$
 MeV.  However, they become briefly overpopulated around $T=160-170$ MeV but then quickly return to the experimental values, except for the case when we have the initial conditions such that the pions are overpopulated.  This could imply that there are a few too many Hagedorn states and a fit for the Hagedorn states with a lower $A$ (degeneracy of the Hagedorn states) may produce better results.  
 
\begin{figure*}
\begin{minipage}{0.45\linewidth}
\centering
\includegraphics[width=3.in]{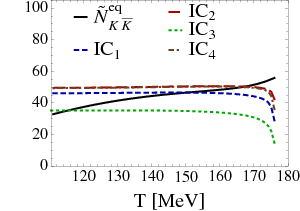}  \\
\includegraphics[width=3.in]{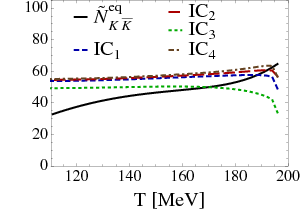} 
\caption{Results for the K's and pions with various initial conditions for $T_H=176$ MeV.} \label{fig:KK176}
\end{minipage}
\hspace{0.5cm}
\begin{minipage}{0.45\linewidth}
\includegraphics[width=3.in]{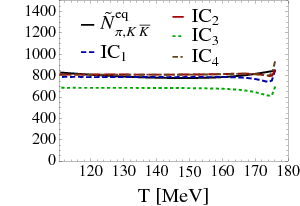}  \\
\includegraphics[width=3.in]{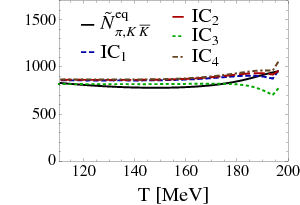}  
\caption{Results for the K's and pions with various initial conditions for $T_H=196$ MeV..} \label{fig:KK196}
\end{minipage}
\end{figure*}

\begin{figure}
\centering
\includegraphics[width=3.in]{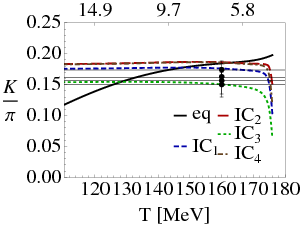}  \\
\includegraphics[width=3.in]{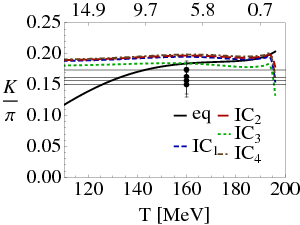} 
\caption{Results for the ratio of K's with various initial conditions. Note that for STAR $K^+\pi^-=0.16$ and  $K^-\pi^-=0.15$.  Along the top axis of each graph the corresponding time is shown in $fm/c$.} \label{fig:KKpi}
\end{figure}

As with the protons, the total number of kaons are also 
slightly dependent on our chosen initial conditions, more specifically, our choice in $\beta_{i}$.  In Fig.\ \ref{fig:KK176} and Fig.\ \ref{fig:KK196} the temperature 
of the evolving system after the phase transition 
at which chemical equilibrium among standard hadrons is basically reached
and maintained
is between $T=160-170$ for $T_H=176$ MeV and they have also already reached chemical equilibrium by $T=170$  for $T_H=196$ MeV,
below which the Hagedorn states basically die out. The one exception is when the Hagedorn states begin underpopulated i.e. that $\beta_{i}<1$. In this case, the kaon pairs take longer to reach chemical equilibrium.  However,  when we look at $K/\pi$ in Fig.\ \ref{fig:KKpi}, lower $\beta_{i}$ actually fits the data better. 

Moreover, the pions again remain roughly at chemical equilibrium throughout the expansion as seen in Fig.\ \ref{fig:KK176} and Fig.\ \ref{fig:KK196} .  While the pion graphs look roughly similar in Figs.\ \ref{fig:pp176}-\ref{fig:KK196}, they are not.  The difference is how the pions are affected in the presence of a $p\bar{p}$ pair compared to a decay that includes a kaon anti-kaon pair.

In Fig.\ \ref{fig:KKpi} the ratio of kaons to pions is shown for $T_H=176$ MeV and for $T_H=196$ MeV.   For $T_H=176$ MeV our results are roughly at the upper edge of the experimental values.  However, for $T_H=196$ MeV our results are slightly higher than the experimental values.  Although, the results at $T=110$ MeV are almost exactly those of the uppermost experimental data point.  

\begin{figure*}
\begin{minipage}{0.45\linewidth}
\centering
\includegraphics[width=3.in]{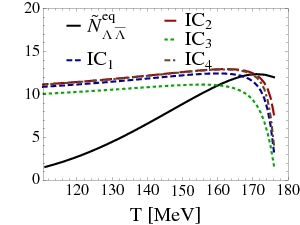}\\
\includegraphics[width=3.in]{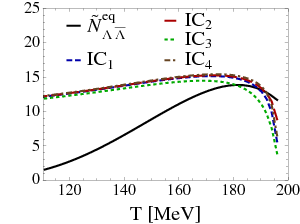} 
\caption{Results for the $\Lambda$'s and pions with various initial conditions for $T_H=176$ MeV.} \label{fig:LL176}
\end{minipage}
\hspace{0.5cm}
\begin{minipage}{0.45\linewidth}
\centering
\includegraphics[width=3.in]{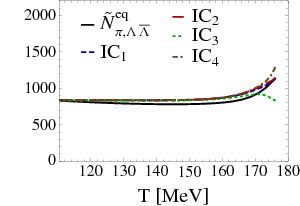} \\
\includegraphics[width=3.in]{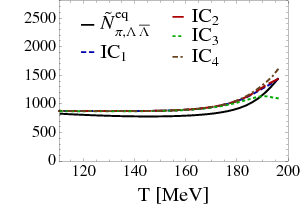}
\caption{Results for the $\Lambda$'s and pions with various initial conditions for $T_H=196$ MeV..} \label{fig:LL196}
\end{minipage}
\end{figure*}

\begin{figure}
\centering
\includegraphics[width=3.in]{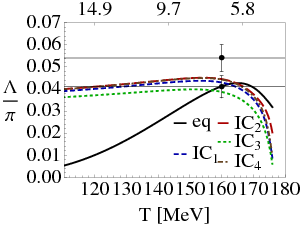}\\
\includegraphics[width=3.in]{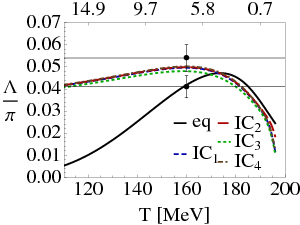}
\caption{Results for the ratio of $\Lambda/\pi$'s with various initial conditions. Note that for STAR $\Lambda/\pi^-=0.54$ and  $\bar{\Lambda}/\pi^-=0.41$.  Along the top axis of each graph the corresponding time is shown in $fm/c$.} \label{fig:LLpi}
\end{figure}

We can also observe the affects of the expansion on the $\Lambda\bar{\Lambda}$ pairs as seen in Fig.\ \ref{fig:LL176} and Fig.\ \ref{fig:LL196}. We see that both reach the experimental values almost immediately ($T>170$ for $T_H=176$ MeV and around $T=190$ for $T_H=196$ MeV). The one exception is again for an underpopulation of Hagedorn states, which reaches chemical equilibrium at $T\approx165$ for $T_H=176$ MeV and already by  $T=170$ for $T_H=196$ MeV).  

The ratio of $\Lambda/\pi$'s is shown in Fig.\ \ref{fig:LLpi}. In both cases the $\Lambda/\pi$'s match the experimental values extremely well.  For $T_H=176$ MeV our results reach the equilibrium values at $T\approx 170$ MeV and for  $T_H=196$ MeV the experimental values are reached already by $T\approx 170$ MeV.

\begin{figure}
\centering
\includegraphics[width=3.in]{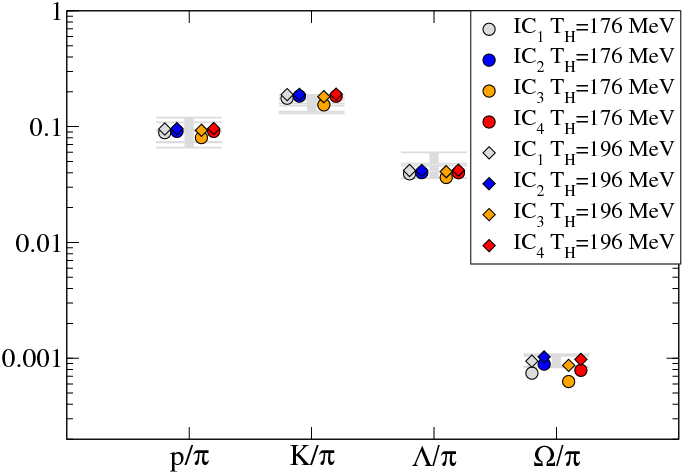}
\caption{Plot of the various ratios including all initial conditions defined in Tab.\ \ref{tab:IC}.  The points show the ratios at $T=110$ MeV for the various initial conditions (circles are for $T_H=176$ MeV and diamonds are for $T_H=196$ MeV).   The experimental results for STAR and PHENIX are shown by the gray error bars.} \label{fig:summary}
\end{figure}

A summary graph of all our results is shown in Fig.\ \ref{fig:summary}.  The gray error bars cover the range of error for the experimental data points from both STAR and PHENIX.  The points show the range in values for the various initial conditions at $T=110$ MeV. We see in our graph that our freezeout results match the experimental data well.

What the graphs in Figs.\ \ref{fig:pp176}-\ref{fig:LLpi} show us is that a dynamical scenario is able to explain chemical equilibration values that appear in thermal fits by $T=160$ MeV. In general, $T_H=176$ MeV and $T_H=196$ give chemical freeze-out values in the range between $T=160-170$ MeV.  These results agree well with the chemical freeze-out temperature found in \cite{NoronhaHostler:2009tz}. 

Moreover, the initial conditions have little effect on the ratios and give a range in the chemical equilibrium temperature of about $\sim5$ MeV, which implies that information from the QGP regarding multiplicities is washed out due to the rapid dynamics of Hagedorn states.  Lower $\beta_i$ does slow the chemical equilibrium time slightly.  However, as seen in Fig.\ \ref{fig:summary} they still fit well within the experimental values. Furthermore, in \cite{Noronha-Hostler:2007fg} we showed the the initial condition play pretty much no roll whatsoever in the ratios of $K/\pi^{+}$ and  $(B+\bar{B})/\pi^{+}$.  Thus, strengthening our argument that the dynamics are washed out following the QGP.

While the variance in the chemical equilibration time  arising from the initial conditions  may seem contradictory to  the $K/\pi^{+}$ and  $(B+\bar{B})/\pi^{+}$  ratios in \cite{Noronha-Hostler:2007fg}, it can be explained with the pion populations.  In Figs.\ \ref{fig:pp176}-\ref{fig:KK196} quicker chemical equilibration times and, thus, larger total baryon/kaon numbers translated into a larger number of pions in the system.  Thus, the $K/\pi^{+}$ and  $(B+\bar{B})/\pi^{+}$  ratios do not depend on the initial conditions.

\section{Production of $\Omega\bar{\Omega}$}\label{omega}

\begin{figure*}
\begin{minipage}{0.45\linewidth}
\centering
\includegraphics[width=3.in]{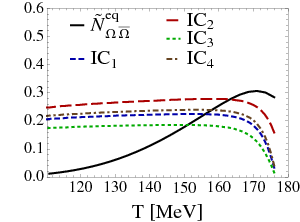} \\
\includegraphics[width=3.in]{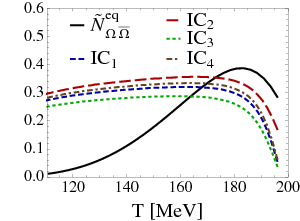}
\caption{Results for the $\Omega$'s and pions with various initial conditions for $T_H=176$ MeV.} \label{fig:OO176}
\end{minipage}
\hspace{0.5cm}
\begin{minipage}{0.45\linewidth}
\centering
\includegraphics[width=3.in]{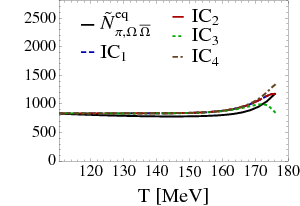} \\
\includegraphics[width=3.in]{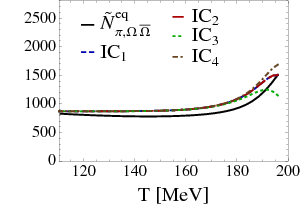}
\caption{Results for the $\Omega$'s and pions with various initial conditions for $T_H=196$ MeV..} \label{fig:OO196}
\end{minipage}
\end{figure*}

\begin{figure}
\centering
\includegraphics[width=3.in]{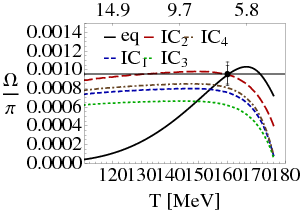} \\
\includegraphics[width=3.in]{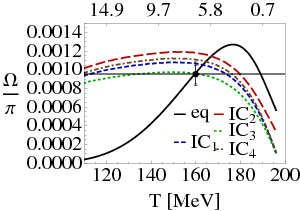}
\caption{Results for the ratio of $\Omega/\pi$'s with various initial conditions. Note that for STAR $\Omega/\pi^-=9.5*10^{-4}$ and  $\bar{\Omega}/\pi^-=9.6*10^{-4}$.  Along the top axis of each graph the corresponding time is shown in $fm/c$.} \label{fig:OOpi}
\end{figure}

We can also use our model to investigate the possibility of $\Omega$'s.  In \cite{Greiner:2004vm}, they discussed the possibility of $\Omega$'s being produced from the following decay channels:
\begin{eqnarray}
HS&\leftrightarrow& \Omega\bar{\Omega} +X\nonumber\\
HS\left(sss\bar{q}\bar{q}\bar{q}\right)&\leftrightarrow&\Omega+\bar{B} +X\nonumber\\
HS_B(sss)&\leftrightarrow&\Omega+X.
\label{Omdecay}
\end{eqnarray}
The first decay channel of a mesonic
non-strange Hagedorn state we can implement
straightforwardly  with our model 
by employing the canonical branching ratio via Fig.\ \ref{fig:lamom}.
The results are shown in Fig.\ \ref{fig:OO176} for $T_H=176$ MeV, in Fig.\ \ref{fig:OO196} for $T_H=196$ MeV, and the $\Omega/\pi$ ratio is shown in Fig.\ \ref{fig:OOpi}.  We are able to find the average number of $\Omega$'s from \cite{ref:max} as seen in Fig.\ \ref{fig:lamom}.   We see that, using only the first reaction,  we are still impressively able to adequately populate the $\Omega\bar{\Omega}$ pairs so that they roughly match the experimental data.  
On the other hand, from Fig.\ \ref{fig:taubb} we see that for the $\Omega $
particle the equilibration time are short only very close to $T_c$.
The scenario is thus more delicate. If one would take eg one half, or one fourth, respectively, 
of the decay width of that of eq. \ref{HSdecaywidth}, the total production
of $\Omega $ is not sufficient up to 25 \%, or up to 50\%, respectively,  to meet the 
experimental yield (the other ratios are not significantly affected by such a change of
the decay width).

In a future work, it would be interesting to observe the other decay channels
as given in Eq. \ref{Omdecay} and advertised in
\cite{Greiner:2004vm}.  The second reaction includes a mesonic,
three times strange Hagedorn state whereas the third decay channel includes a baryonic, strange Hagedorn state. Both states are much more likely
to directly decay into a $\Omega $. 
These are, admittedly, exotic states, but should also occur in the spirit
of Hagedorn states. In order to observe these decay channels 
a method, e.g. a microscopic quark model,  must be found to find the appropriate Hagedorn spectrum for strange mesonic/baryonic Hagedorn states.

\section{Conclusions}\label{conclusions}

In this paper we found that hadronic matter, 
at RHIC or SPS energies, 
%%$K/\pi$
can reach chemical equilibrium within a dynamical scenario using Hagedorn states close to the critical temperature.  These states were able to produce quick chemical equilibration times in (anti-)proton, (anti-)kaons, and (anti-)lambdas
close to the critical temperature due to their strong increase
in their abundancy. The existence of such a mixture of standard hadrons and
Hagedorn states just below the phase transition can explain
dynamically the chemical equilibration of the hadronic species
at around temperatures of 160 MeV to 170 MeV as seen within the thermal models.

 From our analytical results we found that the chemical equilibration time depends on the temperature, decay widths, and branching ratios, but not the initial conditions.  While this changes slightly when an expanding fireball is considered, the initial condition still only play a small role and only minimally affect the 
`freeze-out' temperature at which chemical equilibrium is reached.  This 
demonstrates
 that regardless of the population of hadrons coming out of the QGP phase, the initial conditions are washed out and everything can reach 
 abundances 
 which correspond to those of 
 chemical equilibrium by the chemical freezeout temperatures found in \cite{NoronhaHostler:2009tz}.

Moreover, from our previous paper \cite{Noronha-Hostler:2007fg} we showed that particle ratios ($K/\pi^{+}$ and  $(B+\bar{B})/\pi^{+}$) are not affected by the initial conditions and here we showed that $p/\pi$, $K/\pi$ , $\Lambda/\pi$ 
and also $\Omega / \pi $ match the experimental values regardless of the initial conditions. Specially, Fig.\ \ref{fig:summary} demonstrates this nicely 
and summarizes our findings: Regardless of the initial conditions, our dynamical scenario can match experimental data.  We do find, however, that $T_H=196$ fits within the experimental data box for $K/\pi$ whereas $T_H=176$ is slightly above. This appears to reconfirm the findings in \cite{NoronhaHostler:2009tz}.

Our results imply that both lattice temperature can ensure that the hadrons reach their chemical equilibrium values by $T=160-170$ MeV.  Although the ratios for $T_H=176$ do fit the data somewhat better, both math the experimental values reasonably well.  This implies that independent of the critical temperature the hadrons are able to reach chemical freeze-out.

We see 
sufficiently 
short time scales for the chemical equilibrium of hadrons.  The protons, kaons, and lambdas
reach
chemical equilibrium on the order of 
$\Delta\tau\approx 1-2 \frac{fm}{c}$. Moreover, Hagedorn states states provide a very efficient way for incorporating multi-hadronic
interactions (with parton rearrangements).

In an upcoming paper we will use a canonical model to derive all the branching ratios included in our calculations.  We can then look at reactions that include a mixture of strange and non-strange baryons (for instance, $\bar{p}+\Lambda$) and multi-strange baryons.  However, considering that our initial results produce quick chemical equilibration times for the baryons, kaons, and lambdas, it is reasonable to believe that this will occur for mixed reactions and multi-strange baryons 
as well.   In addition, the machinery of standard hadronic reactions, i.e. binary scattering processes and  resonance production processes, help also to equilibrate the various 
hadronic degrees of freedom.  Still, our work indicates that the population and repopulation of potential
Hagedorn states close to phase boundary 
can be the key source for a dynamical understanding of generating and
chemically equilibrating the standard and measured hadrons.

\section{Acknowledgements}

JNH would like to thank J. Noronha, B. Cole, and M. Gyulassy for productive discussions. 
This work was supported by the Helmholtz International
Center for FAIR within the framework of the
LOEWE program (Landes-Offensive zur Entwicklung
Wissenschaftlich-\"okonomischer Exzellenz) launched by
the State of Hesse.
I.A.S. thanks the members of the Institut f\"ur Theoretische Physik of Johann Wolfgang Goethe--Universit\"at for their hospitality during the final stages of this work. The work of I.A.S. was supported in part by the start-up funds from the Arizona State University.

\begin{appendix}

\section{Appendix: Analytical Solutions
of various equilibartion processes}\label{app}

\begin{figure}
\centering
\includegraphics[width=3.in]{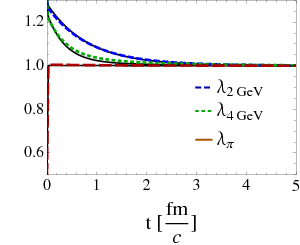}\\
\includegraphics[width=3.in]{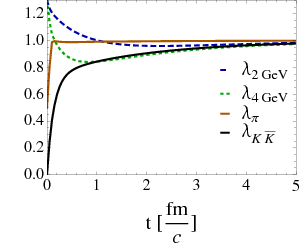}
\caption{Numerical and analytical results for the pions and Hagedorn states where $\eta=0.9$  at $T=175$ MeV  for $T_H=176$ MeV when  $\beta_{i}=1.1$ and $\alpha=0.9$ (top) and the numerical results for the same initial conditions including $K\bar{K}$ pairs with $\phi=0$.} \label{fig:pifree}
\end{figure}

If our initial conditions are such that both the pions and Hagedorn states begin far out of chemical equilibrium, we can find an analytical solution by subdividing the analysis into three distinct stages.  Initially, during stage 1 the pions are underpopulated such that we can say that they approximately begin at $\alpha\approx0$ (we can also start the pions above zero and the approximation works well).  Because the pions reach chemical equilibrium much quicker than the Hagedorn states due to all the Hagedorn states decaying quickly into pions, then we can make the approximation that the Hagedorn states are held at their initial value of $\beta_{i}$.   One can see this from the difference in the time scales from Tab.\ \ref{tab:tau} where $\tau_{i}>\tau_{\pi}^{0}$ and $\tau_{i}>\tau_{\pi}$. Since $\alpha\approx 0$ we let $\lambda_{\pi}^{n}\approx 0$, then substituting this into Eq.\ (\ref{eqn:setpiHS}) we obtain

\begin{eqnarray}\label{eqn:NpiA2}
\dot{\lambda}_{\pi}&=&\sum\Gamma_{i} \frac{N_{i}^{eq}}{N_{\pi}^{eq}}\beta_{i}\langle n_i\rangle ,\nonumber\\
\lambda_{\pi}&=&\left(\frac{t}{\tau_{\pi}^{0}}+\alpha\right)\;,
\end{eqnarray}
which is the fugacity of the pions in stage 1 and gives $\tau_{\pi}^0\equiv\frac{N_{\pi}^{eq}}{\sum_{i} \Gamma_{i} N^{eq}_{i} \langle n_i\rangle\beta_{i}}$.
Again using the approximation $\alpha\approx 0$ and substituting Eq.\ (\ref{eqn:NpiA2}) into the Hagedorn state rate equation in Eq.\ (\ref{eqn:setpiHS}), with the solution
\begin{eqnarray}\label{eqn:s1nipre}
\dot{\lambda_{i}} &=& \Gamma_{i} \left[\left( \frac{t}{\tau_{\pi}^{0}}\right)^{\langle n_i\rangle}-\lambda_{i}\right],\nonumber\\
\lambda_{i}&=& \left[1-\langle n_i\rangle\left(\frac{-t}{\tau_{i}}\right)^{-\langle n_i\rangle} e^{- \left(\frac{t}{\tau_{i}}\right)} \int_{0}^{-\frac{t}{\tau_{i}}}x^{\langle n_i\rangle-1}e^{-x}dx    \right]\nonumber\\
&\cdot& \left(\frac{t}{\tau_{\pi}^{0}}\right)^{\langle n_i\rangle}+\beta_{i}e^{-\left(\frac{t}{\tau_{i}}\right)} .
\end{eqnarray}
Substituting $x=\frac{t}{\tau_{i}}\xi$ into the integral in Eq.\ (\ref{eqn:s1nipre}), expanding the exponential inside the integral so $e^y=\sum_{j=0}^{\infty}\frac{y^j}{j!}$, and integrating over $\xi$, provides us with the fugacity of the Hagedorn states in stage 1
\begin{eqnarray}\label{eqn:HSstage1}
\lambda_{i}&=&\left(\frac{t}{\tau_{\pi}^{0}}\right)^{\langle n_i\rangle} \left[1- e^{- \left(\frac{t}{\tau_{i}}\right)} \sum_{j=0}^{\infty}\frac{\langle n_i\rangle}{j!(\langle n_i\rangle +j)}\left(\frac{t}{\tau_{i}}\right)^j \right]\nonumber\\
&+& \beta_{i}e^{- \left(\frac{t}{\tau_{i}}\right)}\;.
\end{eqnarray}
Therefore, Eq.\ (\ref{eqn:NpiA2}) and Eq.\ (\ref{eqn:HSstage1}) describe the behaviour of the pions and Hagedorn states during the initial stage of the evolution towards chemical equilibrium.  They are then compared to the numerical results in Fig.\ \ref{fig:pifree}.

As the pions near equilibrium our approximation of $\lambda_{\pi}\approx 0$ no longer holds and we switch to stage 2 where we assume $\lambda_{\pi}\approx 1$ at time $t_{1}$. Here $t_{1}$ is a time when the pions are almost in chemical equilibrium, which is normally taken when the pions reach about $\sim 95\%$ of their chemical equilibrium value.  Returning to the pion equation in Eq.\ (\ref{eqn:setpiHS}), we can substitute in $\lambda_{\pi}=1-\epsilon$ and use the approximation $(1-\epsilon)^n\approx1-n\epsilon$
\begin{eqnarray}\label{eqn:HSineqsub}
\dot{\epsilon}&=&-\sum\Gamma_{i} \frac{N_{i}^{eq}}{N_{\pi}^{eq}}\left((\beta_{i}-1)\langle n_i\rangle+ \langle n_i^2\rangle \epsilon\right)\; .  
\end{eqnarray}
Additionally, we substituted in $\beta_i$ for $\lambda_i$ as an approximation since the Hagedorn states do not change significantly in Stage 1 (the majority of the evolution is done by the pions).  Recall that $\beta_i=\lambda_i(t=0)$ and it is a constant.
In its present form, Eq.\ (\ref{eqn:HSineqsub}) can be integrated. We also define $\epsilon(t_{1})=1-\eta$ where $\eta$ is close to 1 ($\eta$ is the measurement of how close the pions are to their equilibrium value when we switch from Stage 1 to Stage 2). Then, after integration
\begin{equation}\label{eqn:Npi}
\lambda_{\pi}=\left(1+\gamma-\left(1+\gamma-\eta\right)e^{-\frac{t-t_{1}}{\tau_{\pi}}}\right)
\end{equation}
where $\gamma=\frac{\sum_{i}\Gamma_{i} N_{i}^{eq}(\beta_{i}-1)\langle n_i\rangle}{\sum_{i}\Gamma_{i}N_{i}^{eq}\langle n_i^2\rangle}$ and $\tau_{\pi}\equiv\frac{N_{\pi}^{eq}}{\sum_{i} \Gamma_{i} N^{eq}_{i} \langle n_{i}^2\rangle}$. 
Analogously to stage 1, we substitute the pion equation, i.e., Eq.\ (\ref{eqn:Npi}) into the Hagedorn resonance equation in Eq.\ (\ref{eqn:setpiHS}) and integrate 
\begin{eqnarray}\label{eqn:Nallep1}
\lambda_{i}&=&\left[d e^{-\frac{t-t_{1}}{\tau_{i}}}+1+\langle n_i\rangle\gamma \right.\nonumber\\
&-&\left.\left(\frac{\tau_{\pi}}{\tau_{\pi}-\tau_{i}}\right) \langle n_i\rangle \left(1+\gamma-\eta\right) e^{-\frac{t-t_{1}}{\tau_{\pi}}}\right]
\end{eqnarray}
where
$d=\omega_{i}-1+\langle n_i\rangle\gamma  +\left(\frac{\tau_{\pi}}{\tau_{\pi}-\tau_{i}}\right) \langle n_i\rangle \left(1+\gamma-\eta\right)$
and $\lambda_{i}(t_{1})=\omega_{i}$. Thus, our equations for the evolution of the pions and Hagedorn states are Eq.\ (\ref{eqn:Npi}) and Eq.\ (\ref{eqn:Nallep1}), respectively.  As with stage 1, the evolution equation for the Hagedorn states is dictated by that of the pions.

Stage 3 i.e. quasi-equilibrium begins once the pions and at least one species of Hagedorn resonances ($\tau_{7GeV}$ is the shortest chemical equilibration time) has surpassed its equilibration time ($\tau_{\pi}$ and $\tau_{i}$, respectively). To understand quasi-equilibrium we must use the effective pion number
\begin{equation}\label{eqn:effpions}
\tilde{N}_{\pi}=N_{\pi}+\sum_{i}N_{i}\langle n_i\rangle\;,
\end{equation}
because we need a variable that can observe the effects of both the pions and resonances.  The effective pion number essentially includes the number of effective pions that each Hagedorn state could decay into.   
Thus, we start by taking the derivative of Eq.\ (\ref{eqn:effpions}) in terms of its fugacity
\begin{eqnarray}
\dot{\tilde{\lambda}}_{\pi}&=&\frac{1}{\tilde{N}_{\pi}^{eq}}\left[N_{\pi}^{eq}\dot{\lambda}_{\pi}+\sum_{i}N_{i}^{eq}\dot{\lambda}_{i}\langle n_i\rangle\right]\nonumber\\
&=&\frac{\sum_{i}\Gamma_{i}N_{i}^{eq}}{\tilde{N}_{\pi}^{eq}}\left[\langle n_i\rangle\sum_{n}B_{i,n}\lambda_{\pi}^n-\sum_{n}B_{i, n}n\lambda_{\pi}^n\right]\;.
\end{eqnarray} 
Once again we make the substitution $\lambda_{\pi}=1-\epsilon$ so that

\begin{eqnarray}\label{eqn:needep}
\dot{\tilde{\epsilon}}&=&-\frac{1}{\tilde{N}_{\pi}^{eq}}\sum_{i}\Gamma_{i}N_{i}^{eq}\sigma_i^2 \epsilon\;.
\end{eqnarray} 
where $\sigma_i^2=\langle n_i^2\rangle-\langle n_i\rangle^2$ in the Gaussian distribution of our branching ratios.
To relate $\epsilon$ and $\tilde{\epsilon}$ we return to Eq.\ (\ref{eqn:effpions}) and separate $\lambda_{i}$ into a sum over the resonances in quasi-equilibrium and one over the ``freely" equilibrating resonances 
\begin{equation}
\tilde{\lambda}_{\pi}=\frac{1}{\tilde{N}_{\pi}^{eq}}\left[N_{\pi}^{eq}\lambda_{\pi}+\sum_{QE}N_{i}^{eq}\langle n_i\rangle\lambda_{i}+\sum_{free}N_{i}^{eq}\langle n_i\rangle\lambda_{i}\right].
\end{equation}
Since the pions reach quasi-equilibrium first, i.e., $\tau_{\pi}<\tau_{i}$ near $T_{c}$, we set the $\pi$ rate equation in Eq.\ (\ref{eqn:setpiHS}) equal to zero, which gives $\lambda_{i}\approx\frac{1}{\langle n_i\rangle}\sum_{n}B_{i, n}n\lambda_{\pi}^n$, so
\begin{eqnarray}\label{eqn:eptil}
\tilde{\lambda}_{\pi}&\approx&1-\frac{\left(N_{\pi}^{eq}+\sum_{QE}N_{i}^{eq}\langle n_i^2\rangle\right)\epsilon}{\tilde{N}_{\pi}^{eq}}\nonumber\\
&-&\frac{\sum_{free}\langle n_i\rangle(N_{i}^{eq}-N_{i}^{eq}\lambda_{i})}{\tilde{N}_{\pi}^{eq}}\;.
\end{eqnarray}
Eq.\ (\ref{eqn:eptil}) then has the form $\tilde{\lambda}_{\pi}\approx 1-\tilde{\epsilon}$ where 
\begin{equation}\label{eqn:e2etil}
\tilde{\epsilon}=\frac{\left(N_{\pi}^{eq}+\sum_{QE}N_{i}^{eq}\langle n_i^2\rangle\right)\epsilon}{\tilde{N}_{\pi}^{eq}}+\frac{\sum_{free}\langle n_i\rangle(N_{i}^{eq}-N_{i}^{eq}\lambda_{i})}{\tilde{N}_{\pi}^{eq}}.
\end{equation}
We can then solve for $\epsilon$ in Eq.\ (\ref{eqn:e2etil}) and substitute $\epsilon$ into Eq.\ (\ref{eqn:needep}), which in turn can be integrated.  This leads us to the solution
\begin{eqnarray}
\tilde{\epsilon}&=&\epsilon_{j} e^{-\frac{t-\tau_{j}}{\tau^{QE}_{\pi}}}+\sum_{free}\langle n_i\rangle N_{i}^{eq}-\frac{\sum_{i}\Gamma_{i}N_{i}^{eq}\sigma_i^2}{\tilde{N}_{\pi}^{eq}\sum_{QE}N_{i}\langle n_i^2\rangle}\nonumber\\
&\cdot&\sum_{free}N_{i}^{eq}\langle n_i\rangle e^{-\frac{t-\tau_{j}}{\tau^{QE}_{\pi}}}\int_{0}^{t}e^{\frac{x-\tau_{j}}{\tau^{QE}_{\pi}}}\lambda_{i}(x)dx\;.
\end{eqnarray}
where $j$ stands for the latest resonance to reach chemical equilibrium at that point in time and 
\begin{equation}
\tau^{QE}_{\pi}\equiv\frac{N_{\pi}^{eq}}{\sum_{i} \Gamma_{i} N^{eq}_{i} \sigma_{i}^2}+\frac{\sum_{QE}N_{i}^{eq}\langle n_{i}^2\rangle}{\sum_{i}\Gamma_{i}N_{i}^{eq}\sigma_{i}^2}
\end{equation}
is the quasi-equilibrium time.
Clearly, once all the Hagedorn states have reached chemical equilibrium than $j$ symbolizes the resonance of $M=2$ GeV, since it is the slowest Hagedorn state to equilibrate.  The sums over ``free" is the sum over the Hagedorn states that have not yet surpassed their respective chemical equilibrium time, $\tau_i$.  Once $\tau_{2GeV}$ is reached those sums equal zero.  Therefore, after $\tau_{2GeV}$ all that remains is 
\begin{equation}\label{eqn:remaine}
\tilde{\epsilon}=\epsilon_{2GeV} e^{-\frac{t-\tau_{2GeV}}{\tau^{QE}_{\pi}}}
\end{equation}
where $\tau^{QE}_{\pi}$ is shown in Tab.\  \ref{tab:tau}.
Finally, we rewrite Eq.\ (\ref{eqn:remaine}) in terms of the pion evolution equation
\begin{equation}\label{eqn:piinqe}
\lambda_{\pi}=1-(1-\kappa)e^{-\frac{t-\tau_{2GeV}}{\tau^{QE}_{\pi}}} 
\end{equation}
where $\kappa=\lambda_{\pi}(\tau_{2GeV})$.

Because the resonance equation depends on the population of the pions we substitute Eq.\ (\ref{eqn:piinqe}) into the Hagedorn resonance rate equation in Eq.\ (\ref{eqn:setpiHS}), assuming the pions are near equilibrium (i.e., we use the approximation  $\lambda=1-\epsilon$ and $(1-\epsilon)^n\approx1-n\epsilon$)
\begin{eqnarray}\label{eqn:NImostinqe}
\lambda_{i}&=&\left(\theta_{i}-1+\frac{\tau^{QE}_{\pi}}{\tau^{QE}_{\pi}-\tau_{i}}\langle n_i\rangle(1-\kappa)\right)e^{-\frac{t-\tau_{2GeV}}{\tau_{i}}}\nonumber\\
&+&1-\frac{\tau^{QE}_{\pi}}{\tau^{QE}_{\pi}-\tau_{i}}\langle n_i\rangle(1-\kappa)e^{-\frac{t-\tau_{j}}{\tau^{QE}_{\pi}}}\;.
\end{eqnarray}
where $\theta_{i}=\lambda_{i}(\tau_{2GeV})$. Thus, for stage 3 the population equations for the pions and the Hagedorn states are Eq.\ (\ref{eqn:piinqe}) and Eq.\ (\ref{eqn:NImostinqe}) so long as $t\geq \tau_{2GeV}$.

Fig.\ \ref{fig:pifree} reveals a remarkable close fit with our numerical results for $T=175$ MeV i.e. $T<T_{H}$. 
Thus, the quasi-chemical equilibrium time, $\tau^{QE}$, depends only on $\Gamma_{i}$, $\langle n_i\rangle$, $\sigma_i^{2}$, and $N^{eq}$, which is temperature dependent, but not on our initial conditions.  As mentioned in the text, though, $\tau^{QE}$ includes many non-linear affects that only occur close to the chemical equilibrium.  Thus, the more appropriate time scale is $\tau_\pi^0$ in order to describe the dynamics. 

We also see from Fig.\ \ref{fig:pifree} that when $K\bar{K}$ pairs are included that the pions and Hagedorn resonances equilibrate in roughly the same amount of time, which implies that our analytical solution can still be approximately applied when $K\bar{K}$ pairs are present.  

\end{appendix}

%%
%%%%%%%%%%%%%%%%%%%%%%%%%%%%%%%%%%%%%%%%%%%%%%%%%%%%%%%%%%%%%%%%%%%%%%%

%%%%%%%%%%%%%%%%%%%%%%%%%%%%%%%%%%%%%%%%%%%%%%%%%%%%%%%%%%%%%%%%%%%%%%%


\begin{thebibliography}{99}
\bibitem{Koch:1986ud}
  P.~Koch, B.~Muller and J.~Rafelski,
  %``Strangeness In Relativistic Heavy Ion Collisions,''
  Phys.\ Rept.\  {\bf 142}, 167 (1986).
  %%CITATION = PRPLC,142,167;%%
%\cite{Stock:1999hm}
\bibitem{Rapp:2000gy}
  R.~Rapp and E.~V.~Shuryak,
  %``Resolving the antibaryon production puzzle in high-energy heavy-ion
  %collisions,''
  Phys.\ Rev.\ Lett.\  {\bf 86} (2001) 2980.
  %%CITATION = PRLTA,86,2980;%%
\bibitem{Greiner}
  C.~Greiner,
  %``Importance of multi-mesonic fusion processes on (strange) antibaryon
  %production,''
  AIP Conf.\ Proc.\  {\bf 644}, 337 (2003);
  Heavy Ion Phys.\  {\bf 14}, 149 (2001);
  %%CITATION = NUCL-TH 0011026;%%
  C.~Greiner and S.~Leupold,
  %``Antihyperon production in relativistic heavy ion collision,''
  J.\ Phys.\ G {\bf 27}, L95 (2001).
\bibitem{Braun-Munzinger}
  P.~Braun-Munzinger {\it et al.} 
  Phys.\ Lett.\ B {\bf 344} (1995) 43;
  Phys.\ Lett.\ B {\bf 365} (1996) 1;
  %%CITATION = NUCL-TH 9508020;%%
  %%  C.~Spieles, H.~Stoecker and C.~Greiner,
  %``Hadron production in relativistic nuclear collisions: Thermal hadron
  %source or hadronizing quark-gluon plasma?,''
  Eur.\ Phys.\ J.\  C {\bf 2}, 351 (1998)
  %%
  P.~Braun-Munzinger, I.~Heppe and J.~Stachel,
  %``Chemical equilibration in Pb + Pb collisions at the SPS,''
  Phys.\ Lett.\ B {\bf 465} (1999) 15.
\bibitem{Kapusta}
  J.~I.~Kapusta and I.~Shovkovy,
  %``Thermal rates for baryon and anti-baryon production,''
  Phys.\ Rev.\ C {\bf 68} (2003) 014901;
  %%CITATION = NUCL-TH 0209075;%%
  J.~I.~Kapusta,
  %``Thermal rates for baryon and anti-baryon production,''
  J.\ Phys.\ G {\bf 30} (2004) S351.
  %%CITATION = JPHGB,G30,S351;%%
%\cite{Huovinen:2003sa}
\bibitem{Huovinen:2003sa}
  P.~Huovinen and J.~I.~Kapusta,
  % ``Rate equation network for baryon production in high energy nuclear
  %collisions,''
  Phys.\ Rev.\ C {\bf 69} (2004) 014902.
  %%CITATION = NUCL-TH 0310051;%%
\bibitem{Stock:1999hm}
  R.~Stock,
  %``The parton to hadron phase transition observed in Pb + Pb collisions at
  %158-GeV per nucleon,''
  Phys.\ Lett.\  B {\bf 456} (1999) 277;
  arXiv:nucl-th/0703050.
  %%CITATION = NUCL-TH/0703050;%%
\bibitem{Heinz:2006ur}
  U.~Heinz and G.~Kestin,
  %``Universal chemical freeze-out as a phase transition signature,''
  arXiv:nucl-th/0612105.
  %%CITATION = NUCL-TH/0612105;%%
\bibitem{BSW}
  P.~Braun-Munzinger, J.~Stachel and C.~Wetterich,
  %``Chemical freeze-out and the QCD phase transition temperature,''
  Phys.\ Lett.\ B {\bf 596} (2004) 61.
  %%CITATION = NUCL-TH 0311005;%%
\bibitem{Greiner:2004vm}
  C.~Greiner {\it et al.}  
  %``Chemical equilibration due to heavy Hagedorn states,''
  J.\ Phys.\ G {\bf 31}, S725 (2005).
  %%CITATION = HEP-PH 0412095;%%
\bibitem{Noronha-Hostler:2007fg}
  J.~Noronha-Hostler, C.~Greiner and I.~A.~Shovkovy,
  %``Fast Equilibration of Hadrons in an Expanding Fireball,''
  Phys.\ Rev.\ Lett.\  {\bf 100}, 252301 (2008).
  %%CITATION = PRLTA,100,252301;%%
\bibitem{NoronhaHostler:2009hp}
  J.~Noronha-Hostler, J.~Noronha, H.~Ahmad, I.~Shovkovy and C.~Greiner,
  %``Chemical Equilibration and Transport Properties of Hadronic Matter near
  %$T_c$,'' 
  arXiv:0907.4963 [nucl-th], to appear in Nucl.\ Phys.\ A; J.~Noronha-Hostler, C.~Greiner and I.~Shovkovy,
  %``Chemical equilibration of baryons in an expanding fireball,''
  Eur.\ Phys.\ J.\ ST {\bf 155}, 61 (2008);
  arXiv:nucl-th/0703079.

%\cite{NoronhaHostler:2008ju}
\bibitem{NoronhaHostler:2008ju}
  J.~Noronha-Hostler, J.~Noronha and C.~Greiner,
  %``Transport Coefficients of Hadronic Matter near $T_c$,''
  arXiv:0811.1571 [nucl-th], to appear in Phys.\ Rev.\ Lett.\ .
  %%CITATION = ARXIV:0811.1571;%%
\bibitem{KSS}
  P.~Kovtun, D.~T.~Son and A.~O.~Starinets,
  %``Viscosity in strongly interacting quantum field theories from black hole
  %physics,''
  Phys.\ Rev.\ Lett.\  {\bf 94}, 111601 (2005).
  %%CITATION = PRLTA,94,111601;%%
%\cite{NoronhaHostler:2008ju}
\bibitem{Kharzeev:2007wb}
  D.~Kharzeev and K.~Tuchin,
  %``Bulk viscosity of QCD matter near the critical temperature,''
  arXiv:0705.4280 [hep-ph].
%\cite{NoronhaHostler:2009tz}
\bibitem{NoronhaHostler:2009tz}
  J.~Noronha-Hostler, H.~Ahmad, J.~Noronha and C.~Greiner,
  %``Particle Ratios as a Probe of the QCD Critical Temperature,''
  arXiv:0906.3960 [nucl-th].
  %%CITATION = ARXIV:0906.3960;%%
\bibitem{zodor}
%%
Y.~Aoki, Z.~Fodor, S.~D.~Katz and K.~K.~Szabo,
 JHEP {\bf 0601}, 089 (2006);
Phys.\ Lett.\  B {\bf 643}, 46 (2006) 
%% 
%\cite{Cheng:2007jq}
\bibitem{Cheng:2007jq}
  M.~Cheng {\it et al.},
  %``The QCD Equation of State with almost Physical Quark Masses,''
  Phys.\ Rev.\  D {\bf 77}, 014511 (2008).
  %%CITATION = PHRVA,D77,014511;%%

%\cite{Bazavov:2009zn}
\bibitem{Bazavov:2009zn}
  A.~Bazavov {\it et al.},
  %``Equation of state and QCD transition at finite temperature,''
  arXiv:0903.4379 [hep-lat].
  %%CITATION = ARXIV:0903.4379;%%

\bibitem{Hagedorn:1968jf}
  R.~Hagedorn,
  % ``Statistical thermodynamics of strong interactions at high energies. 3.
  %Heavy-pair (quark) production rates,''
  Nuovo Cim.\ Suppl.\  {\bf 6}  311 (1968);
  Nuovo Cim.\ Suppl.\  {\bf 3}, 147 (1965).

  %%CITATION = NUCUA,6,311;%%
%\cite{Broniowski:2004yh}
\bibitem{Broniowski:2004yh}
  W.~Broniowski, W.~Florkowski and L.~Y.~Glozman,
  %``Update of the Hagedorn mass spectrum,''
  Phys.\ Rev.\  D {\bf 70}, 117503 (2004)
  [arXiv:hep-ph/0407290].
  %%CITATION = PHRVA,D70,117503;%%
%\cite{Bugaev:2008nu}
\bibitem{Bugaev:2008nu}
  K.~A.~Bugaev, V.~K.~Petrov and G.~M.~Zinovjev,
  %``Why Don't We See the Hagedorn Mass Spectrum in the Experiments?,''
  arXiv:0801.4869 [hep-ph].
  %%CITATION = ARXIV:0801.4869;%%
%\cite{Moretto:2006zz}
\bibitem{Moretto:2006zz}
  L.~G.~Moretto, L.~Phair, K.~A.~Bugaev and J.~B.~Elliott,
  %``The Hagedorn Thermostat,''
  PoS C {\bf POD2006} (2006) 037;
  L.~G.~Moretto, K.~A.~Bugaev, J.~B.~Elliott and L.~Phair,
  %``Can a Hagedorn system have a temperature other than $T_C$ or can a
  %thermostat have a temperature other than its own?,''
  arXiv:nucl-th/0601010;
  arXiv:hep-ph/0511180.
%%
\bibitem{Zakout:2006zj}
  I.~Zakout, C.~Greiner and J.~Schaffner-Bielich,
  %``The order, shape and critical point for the quark-gluon plasma phase
  %transition,''
  Nucl.\ Phys.\  A {\bf 781}, 150 (2007);
  %%CITATION = NUPHA,A781,150;%%;
  I.~Zakout and C.~Greiner,
  %``The thermodynamics for a hadronic gas of fireballs with internal color
  %structures,''
  Phys.\ Rev.\  C {\bf 78}, 034916 (2008);
  %%CITATION = PHRVA,C78,034916;%%\bibitem{Ferroni:2008ej}
  L.~Ferroni and V.~Koch,
  %``Crossover transition in bag-like models,''
  arXiv:0812.1044 [nucl-th].
  %%CITATION = ARXIV:0812.1044;%%
\bibitem{Kapusta:1982qd}
  J.~I.~Kapusta and K.~A.~Olive,
  %``Thermodynamics Of Hadrons: Delimiting The Temperature,''
  Nucl.\ Phys.\  A {\bf 408}, 478 (1983).
  %%CITATION = NUPHA,A408,478;%%
%\cite{Rischke:1991ke}
\bibitem{Rischke:1991ke}
  D.~H.~Rischke, M.~I.~Gorenstein, H.~Stoecker and W.~Greiner,
  %``Excluded Volume Effect For The Nuclear Matter Equation Of State,''
  Z.\ Phys.\  C {\bf 51}, 485 (1991).
  %%CITATION = ZEPYA,C51,485;%%
\bibitem{URQMD}
  S.~A.~Bass {\it et al.},
  %``Microscopic models for ultrarelativistic heavy ion collisions,''
  Prog.\ Part.\ Nucl.\ Phys.\  {\bf 41}, 255 (1998)
  [Prog.\ Part.\ Nucl.\ Phys.\  {\bf 41}, 225 (1998)]
  [arXiv:nucl-th/9803035].
  %%CITATION = PPNPD,41,225;%%
  M.~Bleicher {\it et al.},
  %``Relativistic hadron hadron collisions in the ultra-relativistic quantum
  %molecular dynamics model,''
  J.\ Phys.\ G {\bf 25}, 1859 (1999)
  [arXiv:hep-ph/9909407].
  %%CITATION = JPHGB,G25,1859;%%
  
  

\bibitem{Pal:2005rb}
  S.~Pal and P.~Danielewicz,
  %``Hadron production from resonance decay in relativistic collisions,''
  Phys.\ Lett.\  B {\bf 627}, 55 (2005)
  [arXiv:nucl-th/0505049].
  %%CITATION = PHLTA,B627,55;%%
\bibitem{Liu}
  F.~M.~Liu, K.~Werner and J.~Aichelin,
  %``Comparison of micro-canonical and canonical hadronization,''
  Phys.\ Rev.\ C {\bf 68} (2003) 024905;
  %%CITATION = HEP-PH 0304174;%%
  F.~M.~Liu, et. al.,
   %``A micro-canonical description of hadron production in proton proton
  %collisions,''
  J.\ Phys.\ G {\bf 30} (2004) S589;
  Phys.\ Rev.\ C {\bf 69} (2004) 054002.
\bibitem{Becattini:2004rq}
  F.~Becattini and L.~Ferroni,
  %``Statistical hadronization and hadronic micro-canonical ensemble. II,''
  Eur.\ Phys.\ J.\  C {\bf 38}, 225 (2004)
  %%CITATION = EPHJA,C38,225;%%
\bibitem{ref:max}
  M.~Beitel, J. Noronha-Hostler,  and C.~Greiner,
  Diplomthesis, To appear.
\bibitem{Senda}
  I.~Senda,
  %``A Nucleation Model Of Hadrons Based On QCD String,''
  Phys.\ Lett.\ B {\bf 263}, 270 (1991);
  %%CITATION = PHLTA,B263,270;%%
  F.~Lizzi and I.~Senda,
  %``The Nucleation Model Of Strings And The Hagedorn Phase Transition,''
  Nucl.\ Phys.\ B {\bf 359}, 441 (1991);
  %%CITATION = NUPHA,B359,441;%%
  F.~Lizzi and I.~Senda,
  %``A Model Of Interacting Strings And The Hagedorn Phase Transition,''
  Phys.\ Lett.\ B {\bf 244}, 27 (1990).
  %%CITATION = PHLTA,B244,27;%% 
\bibitem{Eidelman:2004wy}
  S.~Eidelman {\it et al.}  
  Phys.\ Lett.\  B {\bf 592} (2004) 1.
\bibitem{StatModel}
  C.~Spieles, H.~Stoecker and C.~Greiner,
  %``Hadron production in relativistic nuclear collisions: Thermal hadron
  %source or hadronizing quark-gluon plasma?,''
  Eur.\ Phys.\ J.\  C {\bf 2}, 351 (1998);
    C.~Greiner, D.~H.~Rischke, H.~Stoecker and P.~Koch,
  %``The Creation of Strange Quark Matter Droplets as a Unique Signature for
  %Quark - Gluon Plasma Formation in Relativistic Heavy Ion Collisions,''
  Phys.\ Rev.\  D {\bf 38}, 2797 (1988);
  C.~Greiner and H.~Stoecker,
  %``Distillation And Survival Of Strange Quark Matter Droplets In
  %Ultrarelativistic Heavy Ion Collisions,''
  Phys.\ Rev.\  D {\bf 44}, 3517 (1991).
\bibitem{Bearden:2004yx}
  I.~G.~Bearden {\it et al.}  [BRAHMS Collaboration],   
  Phys.\ Rev.\ Lett.\  {\bf 94} (2005) 162301.
\bibitem{Greiner:1993jn}
  C.~Greiner, C.~Gong and B.~Muller,
  %``Some remarks on pion condensation in relativistic heavy ion collisions,''
  Phys.\ Lett.\  B {\bf 316}, 226 (1993).
\bibitem{Becattini:2004rq}
  F.~Becattini and L.~Ferroni,
  %``Statistical hadronization and hadronic microcanonical ensemble. II,''
  Eur.\ Phys.\ J.\  C {\bf 38}, 225 (2004)
  [arXiv:hep-ph/0407117].
  %%CITATION = EPHJA,C38,225;%%

\end{thebibliography}
\end{document}